\documentclass[10pt]{article}
\usepackage{fullpage}

\usepackage{amssymb}
\usepackage{amsmath}
\usepackage{cmll} 
\usepackage{stmaryrd} 
\usepackage{enumitem} 
\usepackage{proof} 
\usepackage{url}
\usepackage[disable]
{todonotes}

\usepackage[normalem]{ ulem }
\usetikzlibrary{calc} %
\usetikzlibrary{arrows}

\usepackage[utf8]{inputenc}

\newtheorem{theorem}{Theorem}

\newtheorem{lemma}[theorem]{Lemma}

\newtheorem{example}[theorem]{Example}

\newcommand{\proofitem}[1]{\paragraph*{\mdseries\textit{#1}}}

\newcommand{\Beginproof}{\proofitem{Proof.}}
\newcommand{\Endproof}{
  \ifmmode 
  \else \leavevmode\unskip\penalty9999 \hbox{}\nobreak\hfill
  \fi
  \quad\hbox{$\Box$}
  \par\medskip}

\newcommand\Eqref[1]{(\ref{#1})}

\renewcommand{\phi}{\varphi}
\renewcommand\epsilon{\varepsilon}

\newcommand{\Implies}{\Rightarrow}

\newcommand{\St}{\mid}

\renewcommand{\Bot}{{\mathord{\perp}}}
\newcommand{\Top}{\top}

\newcommand\cB{\mathcal{B}}

\newcommand\cG{\mathcal{G}}

\newcommand\cM{\mathcal{M}}

\newcommand\cP{\mathcal{P}}

\newcommand\cS{\mathcal{S}}

\newcommand\cU{\mathcal{U}}
\newcommand\cV{\mathcal{V}}

\newcommand\cX{\mathcal{X}}
\newcommand\cY{\mathcal{Y}}

\newcommand\Fini{{\mathrm{fin}}}

\newcommand{\Linarrow}{\multimap}

\newcommand\Myleft{}
\newcommand\Myright{}

\newcommand\Web[1]{\Myleft|{#1}\Myright|}

\newcommand\ITens{\otimes}
\newcommand\Tens[2]{{#1}\ITens{#2}}
\newcommand\Tensp[2]{({#1}\ITens{#2})}
\newcommand\ITensc{\mathbin{\widehat\otimes}}
\newcommand\Tensc[2]{{#1}\ITensc{#2}}
\newcommand\Tenscp[2]{({#1}\ITensc{#2})}
\newcommand\IWith{\mathrel{\&}}
\newcommand\With[2]{{#1}\IWith{#2}}
\newcommand\Withp[2]{({#1}\IWith{#2})}

\newcommand\Orth[2][]{#2^{\Bot_{#1}}}
\newcommand\Orthp[2][]{(#2)^{\Bot_{#1}}}

\newcommand\Pair[2]{\langle{#1},{#2}\rangle}

\newcommand\Biorth[1]{#1^{\Bot\Bot}}

\newcommand\One{1}

\newcommand\Partfin[1]{{\cP_\Fini}({#1})}

\newcommand\Card[1]{\#{#1}}

\newcommand\FamRestr[2]{{#1}|_{#2}}

\newcommand\Locun[1]{1^J}

\newcommand\Isom\simeq

\newcommand\Comp{\mathrel\circ}

\newcommand\Funinv[1]{#1^{-1}}

\newcommand\SET{\mathbf{Set}}

\newcommand\Limpl[2]{{#1}\Linarrow{#2}}
\newcommand\Limplm[2]{{#1}\Linarrow_\Meassymb{#2}}

\newcommand\Limplp[2]{({#1}\Linarrow{#2})}

\newcommand\Meas[1]{\underline{\mathsf{M}}(#1)}
\newcommand\Measm[1]{\mathsf{M}(#1)}

\newcommand\Nat{{\mathbb{N}}}

\newcommand\Snat{\mathsf N}

\newcommand\List[3]{#1_{#2},\dots,#1_{#3}}

\newcommand\Kronecker[2]{\delta_{{#1},{#2}}}

\newcommand\Real{\mathbb{R}}
\newcommand\Realp{\mathbb{R}_{\geq 0}}
\newcommand\Realpto[1]{(\Realp)^{#1}}
\newcommand\Realto[1]{\Real^{#1}}

\newcommand\Dirac[1]{\delta_{#1}}

\newcommand\Evlin{\operatorname{\mathsf{ev}}}

\newcommand\Norm[1]{\|{#1}\|}

\newcommand\Redst[1]{\mathop{\mathsf{Red}}}

\newcommand\Tuple[1]{\langle{#1}\rangle}

\newcommand\Msetofsubst[1]{\bar F}

\newcommand\Inv[1]{{#1}^{-1}}

\newcommand\Dummy{\_}

\newcommand\Pcoh[1]{\mathsf P{#1}}
\newcommand\Pcohp[1]{\mathsf P(#1)}

\newcommand\Base[1]{e_{#1}}
\newcommand\Matapp[2]{{#1}\Compl{#2}}

\newcommand\PCOH{\mathbf{Pcoh}}

\newcommand\Assoc{\alpha}

\newcommand\Symlimpl{\sigma^{\Linarrow}}

\newcommand\Absval[1]{\left|{#1}\right|}

\newcommand\Retri\zeta
\newcommand\Retrp\rho

\newcommand\Impl[2]{[{#1}\rightarrow{#2}]}

\newcommand\Tnat\iota

\newcommand\Loop\Omega

\newcommand\Timpl\Impl
\newcommand\Simpl\Impl

\newcommand\Id{\operatorname{\mathrm{Id}}}

\newcommand\Proj[1]{{\mathsf{pr}}_{#1}}
\newcommand\Inj[1]{{\mathsf{in}}_{#1}}

\newcommand\Excl[1]{\oc{#1}}

\newcommand\Relincl\eta
\newcommand\Relrestr\rho

\newcommand\Compl{\,}

\newcommand\Curlin{\operatorname{\mathsf{cur}}}

\newcommand\Op[1]{{#1}^{\mathsf{op}}}

\newcommand\Eval[2]{\langle#1,#2\rangle}

\newcommand\Obj[1]{\mathsf{Obj}(#1)}

\newcommand\Vect[1]{\overrightarrow{#1}}

\newcommand\Onelem{*}

\newcommand\Cuball[1]{\mathcal B#1}

\newcommand\Cuballp[1]{\mathcal B(#1)}

\newcommand\Meassymb{\mathsf m}

\newcommand\SCCLin{\mathbf{CLin}}
\newcommand\SCCLinm{\SCCLin_\Meassymb}
\newcommand\SCCStab{\mathbf{Cstab}}
\newcommand\SCCStabm{\SCCStab_\Meassymb}
\newcommand\CCLin{\mathbf{CLin}}

\newcommand\Dual[1]{{#1}'}

\newcommand\Cdual[1]{#1'}

\newcommand\Cvx[1]{\operatorname{\mathsf{cvx}}(#1)}

\newcommand\Eset[1]{\{#1\}}

\newcommand\Adj{\dashv}

\newcommand\Wcard{\_{}\,}

\newcommand\Cat[1]{\mathbf{#1}}

\newcommand\ConeofPCS[1]{\overline{\mathsf P}#1}
\newcommand\ConeofLPCS[1]{\overline{\mathsf P}_\infty#1}
\newcommand\ConeofPCSp[1]{\ConeofPCS{(#1)}}

\newcommand\Funofpt[1]{\widehat{#1}}

\newcommand\Interval[2]{[#1,#2]}
\newcommand\Intervaloc[2]{(#1,#2]}

\newcommand\Clinfty{\mathbf L_\infty}

\newcommand\Tonatural{\Rightarrow}

\newcommand\Funcat[2]{[#1,#2]}

\newcommand\Slice[2]{#1/#2}
\newcommand\Slicep[2]{(#1/#2)}
\newcommand\Slicepr[1]{\Delta_{#1}}
\newcommand\Slicec[1]{\gamma^{#1}}

\newcommand\Densext[1]{\widetilde{#1}}

\newcommand\Objc[1]{\Obj{\Cat{#1}}}

\newcommand\Tensciso{\pi}

\newcommand\Assocli{\Assoc^0}
\newcommand\Assocc{\widetilde\Assoc}
\newcommand\Symc{\widetilde\sigma}
\newcommand\Leftuc{\widetilde\lambda}
\newcommand\Rightuc{\widetilde\rho}

\newcommand\Tenscbil{\tau}

\newcommand\Curlinc{\operatorname{\mathsf{cur}}}

\newcommand\CCBilin[3]{\CCLin(#1,#2;#3)}
\newcommand\SCCBilin[3]{\SCCLin(#1,#2;#3)}
\newcommand\Bilunc{\beta}
\newcommand\Linofbilin[1]{\widetilde{#1}}

\newcommand\Bilforms[2]{\cB(#1,#2)}

\newcommand\Parte[1]{\cP^+(#1)}
\newcommand\Parto[1]{\cP^-(#1)}

\newcommand\Staboflin{\mathsf D}
\newcommand\Linofstab{\mathsf E}

\newcommand\Appsume[2]{\Delta^+{#1}(#2)}
\newcommand\Appsumo[2]{\Delta^-{#1}(#2)}

\newcommand\Exclc[1]{\widehat\oc{#1}}
\newcommand\Derc[1]{\operatorname{{\widehat{\mathsf{der}}}}_{#1}}
\newcommand\Diggc[1]{\operatorname{\widehat{{\mathsf{dig}}}}_{#1}}

\newcommand\Seelyc{\widehat{\mathsf m}}
\newcommand\Seelyzc{\Seelyc^0}
\newcommand\Seelytc{\Seelyc^2}

\newcommand\Listadj{\chi}
\newcommand\Promfunc{\widehat{\mathsf{prom}}}
\newcommand\Linfactor[1]{\widetilde{#1}}

\newcommand\Promc[1]{{#1}^{\widehat\oc}}
\newcommand\Prommc[1]{{#1}^{\widehat\oc\widehat\oc}}

\newcommand\MSP{\mathbf{Meas}}
\newcommand\MREFC{\mathbf{M}}
\newcommand\Mref{\mathsf{Mref}}

\newcommand\Mconec[1]{\underline{#1}}
\newcommand\Mconet[1]{\cM(#1)}

\newcommand\Metabs[2]{\boldsymbol\lambda{#1}\,{#2}}

\newcommand\Mpath[1]{\mathsf{paths}_1(#1)}
\newcommand\Mpathu[1]{\mathsf{paths}_1(#1)}
\newcommand\Lftest[2]{#1\mathbin\triangleright #2}

\newcommand\Rzero{0}

\newcommand\Measev[1]{\epsilon_{#1}}
\newcommand\Linofkern[1]{f_{#1}}

\newcommand\Tokern{\leadsto}

\title{On the linear structure of cones}   
\author{Thomas Ehrhard\\
IRIF, CNRS and Univeristy of Paris}

\begin{document}


\maketitle

\begin{abstract}
  For encompassing the limitations of probabilistic coherence spaces
  which do not seem to provide natural interpretations of continuous
  data types such as the real line, Ehrhard and al.~introduced a model
  of probabilistic higher order computation based on (positive) cones,
  and a class of totally monotone functions that they called
  ``stable''. Then Crubillé proved that this model is a
  conservative extension of the earlier probabilistic coherence
  space model. We continue these investigations by showing that the
  category of cones and linear and Scott-continuous functions is a
  model of intuitionistic linear logic. To define the tensor product,
  we use the special adjoint functor theorem, and we prove that this
  operation is and extension of the standard tensor product of
  probabilistic coherence spaces. We also show that these latter are
  dense in cones, thus allowing to lift the main properties of the
  tensor product of probabilistic coherence spaces to general
  cones. Last we define in the same way an exponential of cones and
  extend measurability to these new operations.
\end{abstract}


\section{Introduction}

We continue a series of investigations initiated by Danos and
Ehrhard~\cite{DanosEhrhard08} on a class of models of higher order
computation, based on an initial idea of Girard~\cite{Girard04a}. In
these models, types are interpreted as concrete structures called
\emph{probabilistic coherence spaces} (PCSs) consisting of a set (the
web) and a collection of $\Realp$-valued families indexed by the web
generalizing discrete probability distributions: a typical example of
PCS is $\Nat$ equipped with subprobability\footnote{Not probability, in
  order to interpret also partial computation.}  distributions on
$\Nat$. Another example is $\Nat\times\Nat$ equipped with all families
$(t_{i,j})_{(i,j)\in\Nat\times\Nat}$ such that, for all subprobability
distribution $(x_i)_{i\in\Nat}$ on $\Nat$, the family
$(\sum_{i\in\Nat}t_{i,j}x_i)_{j\in\Nat}$ is a subprobability
distribution on $\Nat$. Such a $t$ is a
$\Nat\times\Nat$ substochastic matrix which represents a sub-Markov
process with $\omega$ states. In~\cite{DanosEhrhard08} it is proven that
PCSs are a categorical model of classical linear logic (LL), that is, a
Seely category (\cite{Mellies09}\footnote{Our main reference for the
  category theory of models of linear logic, see also that paper for
  thorough discussions on the complicated history of the notions
  involved.}) $\PCOH$, where all recursive types can be interpreted,
and which provides an adequate interpretation of a probabilistic
extension of Plotkin's PCF~\cite{Plotkin77}.

\cite{EhrhardPaganiTasson11,EhrhardPaganiTasson14,EhrhardPaganiTasson18,CrubilleEhrhardPaganiTasson17,EhrhardTasson19}
extended these results, proving full abstraction properties for
probabilistic versions of PCF and Levy's Call-by-Push-Value, and
proving that the exponential of PCSs introduced
in~\cite{DanosEhrhard08} is the free one. One essential
feature of this model is that the morphisms of the associated Kleisli
category are extremely regular and can be seen as analytic functions,
some consequences of this fact are presented in
in~\cite{Ehrhard19a} and crucially used in proofs of full abstraction.

The main weakness of the PCS model is that it does not host
``continuous data types'' such as the real line equipped with its
standard Borel $\Sigma$-algebra, required for taking into account
modern probabilistic languages used in Bayesian programming.
\cite{DanosEhrhard08}~suggested that PCSs might be generalized using a
well-suited notion of ordered Banach space or positive cone. This was
done in~\cite{EhrhardPaganiTasson18b}, using a notion of positive
cone\footnote{There is a long tradition of research on this kind of
  structures, rooted in the theory of Banach spaces. Such cones have
  been used in semantics quite successfully for instance
  in~\cite{KeimelPlotkin17} and subsequent work.} considered earlier
in particular in~\cite{Selinger04}. Any PCS gives rise naturally to
such a cone, and one can also associate with any measurable space the
cone of all measures which have a finite global weight.
\cite{EhrhardPaganiTasson18b}~shows that, equipped with suitable
\emph{stable} morphisms (which are Scott-continuous functions
satisfying a total monotonicity requirement which has some
similarities with Berry's stability), these objects form a cartesian
closed category (CCC) $\SCCStab$ providing an adequate interpretation
of an extension of PCF with a type of real numbers and a sampling
primitive. Then, Crubillé showed that this CCC contains the Kleisli
category of the PCS model as a full sub-CCC~\cite{Crubille18},
providing a very satisfactory connection between these constructions.

Following~\cite{Selinger04}, it is noticed
in~\cite{EhrhardPaganiTasson18b} that there is a natural notion of
linear and Scott-continuous functions between cones, which coincides
with the notion of linear morphisms of $\PCOH$ when restricted to
cones induced by PCSs: this defines the category $\SCCLin$ we study
here. Given cones $P$ and $Q$, one can build a
cone $\Limpl PQ$ whose elements are those of $\SCCLin(P,Q)$ so we
could reasonably expect the functor $\Limpl P\Wcard$ to have a left
adjoint for each $P$, hopefully turning $\SCCLin$ into a symmetric
monoidal closed category (SMCC)\footnote{Probably not a *-autonomous
  category however.}.

With cones $P$ and $Q$ we should associate functorially a cone
$\Tens PQ$ such that (at least) there is a natural bijection between
$\SCCLin(\Tens PQ,R)$ and $\SCCLin(P,\Limpl QR)$. Our first attempt
was concrete: since the elements of this second hom-set are continuous
and bilinear functions $P\times Q\to R$, our tensor product should
classify such functions and hence it was natural to look for
$\Tens PQ$ as a sub-cone\footnote{A notion to be defined carefully.}
of $\Cdual{\Bilforms PQ}$ where $\Bilforms PQ$ is the cone of
continuous bilinear maps $P\times Q\to\Realp$ and
$\Cdual R=\Limplp R\Realp$ (the dual of $R$): with any $x\in P$ and
$y\in Q$ we can indeed associate the linear and continuous function
$\Tens xy:\Bilforms PQ\to\Realp$, $f\mapsto f(x,y)$. Whence a
definition of $\Tens PQ$: the least subcone of $\Cdual{\Bilforms PQ}$
which contains all the $\Tens xy$, for $x\in P$ and $y\in Q$. This
also gives us a continuous and bilinear map
$\tau:P\times Q\to\Tens PQ$,  $(x,y)\mapsto\Tens xy$.

We should now prove the universal property: for any bilinear and
continuous $f:P\times Q\to R$, there is exactly one
$\Linofbilin f\in\SCCLin(\Tens PQ,R)$ such that
$f=\Linofbilin f\Compl\tau$. It is easy to define $\Linofbilin f$ on
the elements of $\Tens PQ$ of shape $\Tens xy$ (under a mild
separateness assumption on our cones), but how can we extend this map
to the whole of $\Tens PQ$? Our ``top-down'' definition of $\Tens PQ$
is ineffective for this, we need a ``bottom-up'' approach, something
like: an element of $\Tens PQ$ is a (possibly infinite) linear
combination $\sum_{i\in I}\alpha_i(\Tens{x(i)}{y(i)})$ where
$\alpha_i\in\Realp$ and $\sum_{i\in I}\alpha_i=1$ (convex combination
of pure tensors). But this is not enough because we could perfectly
have two convex combinations of pure tensors $z$ and $z'$ such that
$z'\leq z$ (in $\Cdual{\Bilforms PQ}$) and then $\Tens PQ$ will also
contain $z-z'$ (Example~\ref{ex:tensor-diff} shows that such
subtractions are mandatory at least if we want our $\ITens$ to extensd
that of $\PCOH$). In the usual algebraic case, coefficients form
a ring and such elements are just combinations of pure tensors, with
possibly negative coefficients. Here on the contrary we have to take
such differences into account explicitly since our coefficients are in
$\Realp$.

Another problem arises from the very peculiar
completeness of cones and continuity of morphisms, which are defined
purely in terms of the algebraic order relation
(according to which $x_1\leq x_2$ if there exists $x$ such that
$x_1+x=x_2$), and not of the norm: if a given element $z$ of
$\Tens PQ$ can be written in two different ways as a convex
combination of pure tensors
$z=\sum_{i\in I}\alpha_i\Tensp{x(i)}{y(i)}=\sum_{j\in
  J}\beta_j\Tensp{x'(j)}{y'(j)}$, it is no obvious, though certainly
true, that
$\sum_{i\in I}\alpha_i f(x(i),y(i))=\sum_{j\in J}\beta_j
f(x'(j),y'(j))$.

\paragraph*{Contents.}
After several attempts, we arrived to the conclusion that the concrete
approach would lead to rather complicated (though quite interesting)
developments. Fortunately a shorter road was open, based on the
following observation: our category $\SCCLin$ is small complete and the
functor $\Limpl P\Wcard$ preserves all small limits so we are in position of
applying the special adjoint functor theorem (because $\SCCLin$ is
also well-powered, and, under the aforementioned separateness
condition on objects, it admits $\Realp$ as cogenerating object). So
the functor $\Limpl P\Wcard$ has a left adjoint: we get our tensor
product $\ITens$ almost for free! This is not the end of the story
however because the simple fact that $\ITens$ is a bifunctor defined
as a left adjoint to $\Limpl{}{}$ is not sufficient to prove that it
defines a monoidal structure. Though, we are lucky again because
\begin{itemize}
\item it turns out that $\PCOH$ is a \emph{dense} subcategory of
  $\CCLin$ (that is, any cone is a colimit of a diagram of PCSs),
  which \emph{per se} is quite an interesting property;
\item being a left adjoint, $\ITens$ commutes with all existing
  colimits of cones;
\item restricted to $\PCOH$, our new tensor product coincides with the
  ordinary one, which defines a monoidal structure
  on $\PCOH$.
\end{itemize}
Combining these facts we lift the monoidal structure (associativity
isomorphisms etc) from $\PCOH$ to $\SCCLin$, thus proving that
$\SCCLin$ is an SMCC, which contains $\PCOH$ as a full sub-SMCC and
this was our main goal.

Then we use the same method to define an exponential functor
$\Excl\Wcard:\SCCLin\to\SCCLin$ and show that it is a resource
modality in the sense of Seely categories (again, see~\cite{Mellies09}).

We conclude the paper by explaining shortly how the
\emph{measurability structure} introduced for cones
in~\cite{EhrhardPaganiTasson18b} can be extended to our tensor product
and exponential. Such structures are indeed essential for interpreting
the sampling constructs of probabilistic programming languages.

\paragraph*{Related work.} Positive cones have been used in various
contexts in the semantics of probabilistic programming languages,
notably under the name of \emph{Kegelspitzen} (which are ``unit
balls'' of cones) for which we refer
to~\cite{KeimelPlotkin17,GoubaultL15}. The main difference with our
approach is that such cones are usually equipped with an additional
``extensional'' order relation whereas the only order relation we
consider in our work is the algebraic one: this constraint, strongly
suggested by PCSs, \emph{obliged} the authors
of~\cite{EhrhardPaganiTasson18b} to introduce stable functions.

Closer to our approach are~\cite{Slavnov19}
and~\cite{DahlquistKozen19} where types are interpreted as ordered
Banach spaces and tensor products are also defined. The main
difference that we can see between their approaches and ours is that
they put more standard continuity requirements on linear morphisms,
based on the norm, whereas we insist on our linear (and stable)
morphisms to be \emph{Scott continuous}, a purely\footnote{Not
  completely actually, since we require commutation with lubs of
  \emph{bounded} monotone sequences, and the definition of boundedness
  involves the norm.}  order-theoretic notion which implies
boundedness and thus norm-based continuity, but the converse
implication does not hold in general. The main benefit of insisting on
this kind of continuity is that, our stable morphisms being
Scott-continuous, they have least fixed points (and by cartesian
closeness, the function computing these fixed points is itself
stable). Deeply related with this choice is the fact that stable
functions are defined only on the unit ball of the source cone: the
use of fixed points prevents in general stable functions from being
extended to the whole cone, see~\cite{Ehrhard19a} for examples
illustrating this fact.

Many proofs are omitted, they can be found in an Appendix.

\section{Density}\label{sec:gen-dense-functors}
The categorical notion of density (see~\cite{MacLane71}, Chap.~X
Sec.~6) plays a crucial role, we spend some time for introducing it
and present useful properties\footnote{There is no doubt that they are
  all quite standard, we provide the statements in a form convenient
  for our purpose, and the proofs for self-containedness.}. But we
start with the following simple lemma will be quite useful.
\begin{lemma}\label{lemma:functor-yoneda-iso}
  Let $\Cat C$ and $\Cat D$ be categories, $F,G:\Cat C\to\Cat D$ be
  functors and $\psi_{C,D}:\Cat D(F(C),D)\to\Cat C(G(C),D)$ be a
  natural bijection. Then the family of morphisms
  $\eta_C=\psi_{C,F(C)}(\Id_{F(C)})\in\Cat D(G(C),F(C))$ is a natural
  isomorphism whose inverse it the family of morphisms
  $\theta_C=\Funinv{\psi_{C,G(C)}}(\Id_{G(C)})\in\Cat D(F(C),G(C))$.
\end{lemma}

A functor $F:\Cat C\to\Cat D$ is cocontinuous if it preserves all
small colimits which exist in $\Cat C$: given a
functor $\Delta:J\to\Cat C$ where $J$ is small (one says that $\Delta$
is a diagram) and given a colimiting cocone
$\gamma:\Delta\Tonatural c$ on $\Delta$ (initial object in the
category of cocones on $\Delta$) for some object $c$ of $\Cat C$, then
the cocone $F\gamma:F\Delta\Tonatural F(c)$ is a colimiting cocone in $\Cat D$.

Given categories $\Cat D$ and $\Cat E$, we use
$\Funcat{\Cat D}{\Cat E}$ for the category of functors and natural
transformations from $\Cat D$ to $\Cat E$.

\begin{lemma}\label{lemma:functor-transpose-cocont}
  Let $F:\Cat C\times\Cat D\to\Cat E$ be a functor which is
  cocontinuous in its first argument (that is, given any object $d$ of
  $\Cat D$, the functor $F(\Dummy, d)$ is cocontinuous). Then the
  transpose functor $F':\Cat C\to\Funcat{\Cat D}{\Cat E}$ is
  cocontinuous.
\end{lemma}


Let $I:\Cat C^0\to \Cat C$ (where we assume $\Cat C^0$ to be small)
and let $c\in\Obj{\Cat C}$. Let $\Slice Ic$ be the comma category (its
objects are the pairs $(x,f)$ where $x\in\Obj{\Cat C^0}$ and
$f\in\Cat C(I(x),c)$ and $\Slice Ic((x,f),(y,g))$ is the set of all
$t\in\Cat C^0(x,y)$ such that $g\Compl I(t)=f$) and
$\Slicepr c:\Slice Ic\to\Cat C$ be the functor which maps $(x,f)$ to
$I(x)$ and similarly for morphisms. Let
$\Slicec c:\Slicepr c\Tonatural c$ be the cocone defined by
$\Slicec c_{(x,f)}=f$. One says that the functor $I$ is dense
(see~\cite{MacLane71}, Chap.~X, Sec.~6) if $\Slicec c$ is a colimiting
cocone for each $c\in\Obj C$. If $\Cat C^0$ is a full
subcategory of $\Cat C$ and $I$ is the inclusion, $\Cat C^0$ is said
to be a dense subcategory of $\Cat C$.

\begin{lemma}\label{lemma:dense-unique-morphism}
  Let $I:\Cat C^0\to\Cat C$ be dense and let $F:\Cat C\to\Cat D$ be
  cocontinuous. Let $c\in\Objc C$, $d\in\Objc D$ and
  $l,l'\in\Cat D(F(c),d)$. If, for all $x\in\Cat C^0$ and
  $f\in\Cat C(I(x),c)$ one has $l\Compl F(f)=l'\Compl F(f)$ then
  $l=l'$.
\end{lemma}
\Beginproof
By our assumption on $l$ and $l'$ we define in $\Cat D$ a unique
cocone $\delta:F\Compl\Slicepr c\Tonatural d$ by setting
$\delta_{(x,f)}=l\Compl F(f)=l'\Compl F(f)$ and the fact that
$F\Slicec c$ is a colimiting cocone (because $F$ is cocontinuous)
implies that $l=l'$.
\Endproof

\begin{lemma}\label{lemma:dense-nat-trans-extension}
  Let $I:\Cat C^0\to\Cat C$ be dense, let $F,G:\Cat C\to\Cat D$ be
  functors and assume that $F$ is cocontinuous. Let
  $\tau:F\Compl I\Tonatural G\Compl I$, there is exactly one
  $\Densext\tau:F\Tonatural G$ such that $\Densext\tau\Compl
  I=\tau$. Moreover if $\tau$ is a natural isomorphism and $G$ is also
  cocontinuous, then $\Densext\tau$ is an isomorphism.
\end{lemma}

Now we extend the previous results to separately cocontinuous
multi-ary functors since we want to apply them to our tensor product.

\begin{lemma}\label{lemma:dense-unique-morphism-multi}
  For $i=1,\dots,n$ let $I_i:\Cat C_i^0\to\Cat C_i$ be dense
  functors. Let $F:\prod_{i=1}^n\Cat C_i\to\Cat D$ be \emph{separately
    cocontinuous} (that is, for each $i=1,\dots,n$ and each
  $c_1\in\Obj{\Cat C_1},\dots,c_{i-1}\in\Obj{\Cat
    C_{i-1}},c_{i+1}\in\Obj{\Cat C_{i+1}},\dots,c_n\in\Obj{\Cat C_n}$,
  the functor $F(\List c1{i-1},\Dummy,\List c{i+1}n)$ is
  cocontinuous). Let $\Vect c\in\Obj{\prod_{i=1}^n\Cat C_i}$,
  $d\in\Obj D$ and let $l,l'\in\Cat D(F(\Vect c),d)$. If, for all
  $\Vect x\in\Obj{\prod_{i=1}^n\Cat C^0_i}$ and all
  $\Vect f\in\prod_{i=1}^n\Cat C_i(I_i(x_i),c_i)$ one has
  $l\Compl F(\Vect f)=l'\Compl F(\Vect f)$, then $l=l'$.
\end{lemma}

\begin{theorem}\label{th:dense-concont-tnat-extension}
  For $i=1,\dots,n$ let $I_i:\Cat C_i^0\to\Cat C_i$ be dense
  functors. Let $F,G:\prod_{i=1}^n\Cat C_i\to\Cat D$ be functors and
  assume that $F$ is \emph{separately cocontinuous}. For any natural
  transformation
  $\tau:F\Compl(\prod_{i=1}^nI_i)\Tonatural
  G\Compl(\prod_{i=1}^nI_i)$, there is exactly one natural
  transformation $\Densext\tau:F\Tonatural G$ such that
  $\Densext\tau\Compl(\prod_{i=1}^nI_i)=\tau$. If $G$ is also
  separately cocontinuous and if $\tau$ is a natural bijection, then
  $\Densext\tau$ is also a natural bijection.
\end{theorem}


\section{The category of cones and linear maps}\label{sec:cones-lin-cont}

A \emph{positive cone} is a structure $(P,\Norm\Wcard)$ where $P$ is an
$\Realp$-semimodule and $\Norm\Wcard$ is a function $P\to\Realp$ which
satisfies the usual conditions of a norm\footnote{It is essential to
  notice that this norm is part of the structure of the cone.}. It is
assumed moreover that $P$ is cancellative (meaning
$x_1+x=x_2+x\Implies x_1=x_2$) and that $(P,\Norm\Wcard)$ is positive
(meaning $\Norm{x_1}\leq\Norm{x_1+x_2}$). A subset $C$ of $P$ is
bounded if $\Eset{\Norm x\St x\in C}$ is bounded in $\Realp$.  We use
$\Cuball P$ for the closed unit ``ball''
$\Eset{x\in P\St\Norm x\leq 1}$. The algebraic order relation of $P$
is defined by: $x_1\leq x_2$ if $\exists x\in P\ x_1+x=x_2$. When such
an $x$ exists it is unique by cancellativity, and we use the notation
$x=x_2-x_1$; apart from its partiality, this
subtraction obeys all the usual algebraic laws. One
says that $P$ is complete if any monotone
$\omega$-indexed\footnote{And not arbitrary directed sets as a
  domain-theorist might expect, because we need to apply the monotone
  convergence theorem of measure theory at some point.}  sequence in
$\Cuball P$ has a lub which lies in $\Cuball P$.

The semiring $\Realp$ is a complete positive cone, with norm defined
as the identity.

\begin{example}
  Given a measurable space $\cX$ (with $\Sigma$-algebra $\Sigma_\cX$),
  the set $\Meas\cX$ of all non-negative measures $\mu$ on $\cX$ such
  that $\mu(\cX)<\infty$ is a complete positive cone, when equipped
  with algebraic operations defined pointwise and norm
  $\Norm\mu=\mu(\cX)$.
\end{example}

\paragraph{Continuity and linearity.}
When dealing with cones, the word ``continuity'' always applies to
functions which are monotone wrt.~the algebraic order, and means
commutation with lubs of such monotone sequences in the unit ball. It
is easy to check that all the operations of a cone (addition, scalar
multiplication and norm) are monotone and continuous.

Given $P$ and $Q$ complete positive cones, a map $f:P\to Q$ is linear
if it commutes with the algebraic operations. If $f$ is moreover
continuous then it is not hard to prove that it is bounded in the
sense that it maps $\Cuball P$ to a bounded subset of $Q$
(\cite{Selinger04}). Therefore we can define
$\Norm f=\sup_{x\in\Cuball P}\Norm{f(x)}\in\Realp$. We use $\Cdual P$
for the set of linear and continuous maps $P\to\Realp$.  We say that
$P$ is \emph{separated}\footnote{It is not completely clear to us that
  all cones are separated as one would expect with 
  Banach spaces in mind.} if
$(\forall x'\in\Cdual P\ \Eval{x_1}{x'}=\Eval{x_2}{x'})\Implies x_1=x_2$.

\paragraph{Infinite sums.}
Let $P$ be a cone and $\Vect x=(x_i)_{i\in I}$ be a family of elements
of $P$ indexed by a set $I$ which is at most countable. We say that
$\Vect x$ is summable if the family of real numbers
$(\Norm{\sum_{i\in J}x_i})_{J\in\Partfin I}$ is bounded. In that case
one can define $\sum_{i\in I}x_i\in P$ in an unique way. Indeed, take
a monotone sequence $I(n)$ of finite subsets of $I$ such that
$\bigcup_{n\in\Nat}I(n)=I$, then the sequence
$(\sum_{i\in I(n)}x_i)_{n\in\Nat}$ is monotone and norm-bounded and
hence has a lub in $P$. This lub does not depend on the choice of the
sequence $(I(n))_{n\in\Nat}$ because any two such sequences are
cofinal. We use $\sum_{i\in I}x_i$ for this lub. Obviously any
sub-family of a summable family is summable.

\begin{lemma}
  Let $f:P\to Q$ be linear and continuous. Then for any summable
  family $(x_i)_{i\in I}$ in $P$, the family $(f(x_i))_{i\in I}$ is
  summable in $Q$ and we have
  $f(\sum_{i\in I}x_i)=\sum_{i\in I}f(x_i)$.
\end{lemma}

\begin{lemma}\label{lemma:commutation-cone-sums}
  Let $\Vect x=(x_{i,j})_{(i,j)\in I\times J}$ be a doubly-indexed
  family of elements of a cone $P$ and assume that for each $i\in I$
  the family $(x_{i,j})_{j\in J}$ is summable and that the family
  $(\sum_{j\in J}x_{i,j})_{i\in I}$ is summable. Then the family
  $\Vect x$ is summable and
  $\sum_{i\in I,j\in J}x_{i,j}=\sum_{i\in I}\sum_{j\in
    J}x_{i,j}=\sum_{j\in J}\sum_{i\in I}x_{i,j}$.
\end{lemma}


We use $\SCCLin$ for the category whose objects are the separated
complete positive cones and morphisms are the continuous linear
functions whose norm is $\leq 1$, in other words, the $f:P\to Q$ which
are linear and continuous and satisfy
$f(\Cuball P)\subseteq\Cuball Q$.

\subsection{Linear function spaces}
Let $P$ be and $Q$ be object of $\SCCLin$, we define the cone
$\Limpl PQ$ whose elements are the linear and continuous functions
$P\to Q$ with algebraic operations defined pointwise and norm defined
by $\Norm f=\sup_{x\in\Cuball P}\Norm{f(x)}_Q$ which is well-defined
by continuity of $f$. Notice that in this cone, the algebraic order
relation coincides with the pointwise order on functions. Let indeed
$f,g\in\Limpl PQ$ be such that $\forall x\in P\ f(x)\leq g(x)$. Then
we define a linear function $h:P\to Q$ by setting $h(x)=g(x)-f(x)$ by
the usual laws satisfied by subtraction. Let us prove that this linear
function $h$ is continuous so let $(x_n)_{n\in\Nat}$ be a
non-decreasing sequence in $\Cuball P$ and let $x\in\Cuball P$ be its
lub, we need to prove that $h(x)\leq\sup_{n\in\Nat}h(x_n)$, the
converse resulting from the monotonicity of $h$, that is, we have to
prove that $g(x)\leq f(x)+\sup_{n\in\Nat}h(x_n)$. Let $k\in\Nat$, one
has $g(x_k)=f(x_k)+h(x_k)\leq f(x)+\sup_{n\in\Nat}h(x_n)$ and we are
done since $g$ is continuous.

The cone $\Limpl PQ$ is complete, lubs being computed pointwise (since
the order relation is the pointwise order on functions). This cone is
separated because, given $f_1,f_2\in\Limpl PQ$ with $f_1\not=f_2$
there exists $x\in P$ such that $f_1(x)\not=f_2(x)$ and hence there
exists $y'\in\Dual Q$ which separates $f_1(x)$ from $f_2(x)$. Now the
operation $h\mapsto\Eval{h(x)}{y'}$ is an element of
$\Cdual{\Limplp PQ}$ which separates $f_1$ from $f_2$.

Moreover the operation $\Limpl\Wcard\Wcard$ is a functor
$\Op\SCCLin\times\SCCLin\to\SCCLin$, the action of morphisms being
defined as follows. Let $f\in\SCCLin(P_2,P_1)$ and
$g\in\SCCLin(Q_1,Q_2)$, then
$\Limpl fg\in\SCCLin(\Limplp{P_1}{Q_1},\Limplp{P_2}{Q_2})$ is given by
$\Limplp fg(h)=g\Compl h\Compl f$. The fact that $\Limpl fg$ is a well
defined linear function $\Limplp{P_1}{Q_1}\to\Limplp{P_2}{Q_2}$
results from the linearity of $f$ and $g$. The fact that it is
continuous results from the fact that the lubs in $\Limpl{P_i}{Q_i}$
are computed pointwise and from the continuity of $g$. The fact that
$\Norm{\Limpl fg}\leq 1$ results from the fact that the norms of $f$
and $g$ are $\leq 1$.

\paragraph{Bilinear maps.}\label{sec:bilinear}
Given cones $P$, $Q$ and $R$, a function $f:P\times Q\to R$ is
bilinear and separately continuous if for all $x\in\Cuball P$ and
$y\in\Cuball Q$, one has $f(\Wcard,y)\in\SCCLin(P,R)$ and
$f(x,\Wcard)\in\SCCLin(Q,R)$.  We use $\CCBilin PQR$ for the set of
these bilinear and separately continuous functions\footnote{Actually
  separate continuity is equivalent to continuity on $P\times Q$
  because our notion of continuity is defined as preservation of lubs
  of bounded monotone families.}.

\begin{lemma}\label{lemma:cone-bilin-limpl}
  There is a natural bijection
  $\Bilunc:\SCCLin(P,\Limpl QR)\Tonatural\CCBilin PQR$ of functors
  $\Op\SCCLin\times\Op\SCCLin\times\SCCLin\to\SET$.
\end{lemma}
\Beginproof
Let $g\in\SCCLin(P,\Limpl QR)$, we define $f:P\times Q\to R$ by
$f(x,y)=g(x)(y)$. It is clear that $f$ is separately linear (that is
the maps $f(\Wcard,y)$ and $f(x,\Wcard)$ are linear) because the
algebraic operations of $\Limpl QR$ are defined pointwise, let us
prove separate continuity. Let $(x(n))_{n\in\Nat}$ be monotone in
$\Cuball P$ and $y\in\Cuball Q$. Then
$f(\sup_{n\in\Nat}x(n),y)=g(\sup_{n\in\Nat}x(n))(y)=\sup_{n\in\Nat}f(x(n),y)$
because lubs of linear functions are computed pointwise in
$\Limpl QR$. Let $x\in\Cuball P$ and $(y(n))_{n\in\Nat}$ be monotone in
$\Cuball Q$, we have
\[
  f(x,\sup_{n\in\Nat}y(n))=g(x)(\sup_{n\in\Nat}y(n))=\sup_{n\in\Nat}f(x,y(n))
\]
since the linear function $g(x)$ is continuous, hence
$f\in\CCBilin PQR$, we set $\Bilunc(g)=f$. Let now
$f\in\CCBilin PQR$. Let $x\in\Cuball P$, then we set
$g(x)=f(x,\Wcard)\in\Limpl QR$. Linearity and continuity of $g$
follow again from the fact that all the operations of $\Limpl QR$
(including lubs) are defined pointwise. Let
$g=\beta'(f)\in\SCCLin(P,\Limpl QR)$. It is clear that $\beta$ and
$\beta'$ are natural and inverse of each other.
\Endproof


\subsection{Probabilistic coherence spaces}
Let $I$ be a set (that we can assume to be at most countable). Given
$u,u'\in\Realpto I$, we set
$\Eval u{u'}=\sum_{i\in I}u_iu'_i\in\Realp\cup\Eset\infty$. Given
$\cU\subseteq\Realpto I$, we set
$\Orth\cU=\Eset{u'\in\Realpto I\St\forall u\in\cU\ \Eval u{u'}\leq
  1}$. A \emph{probabilistic coherence space} (PCS) is a pair
$X=(\Web X,\Pcoh X)$ where $\Web X$ is a set (which can be assumed at
most countable) and $\Pcoh X\subseteq\Realpto{\Web X}$ such that
$\Pcoh X=\Biorth{\Pcoh X}$ and
$\forall a\in\Web X\ 0<\sup\Eset{u_a\St u\in\Pcoh X}<\infty$, the
purpose of this second condition being of keeping all coefficients
finite. We set
$\ConeofPCS X=\Eset{u\in\Realpto{\Web X}\St\exists\lambda>0\ \lambda
  u\in\Pcoh X}$. Equipped with algebraic operations defined pointwise,
it is a cancellative $\Realp$-semimodule. We define a norm by setting
$\Norm u=\sup\Eset{\Eval u{u'}\St u'\in\Orth{\Pcoh X}}$ and it is
easily checked that this turns $\ConeofPCS X$ into a separated
complete cone such that $\Cuballp{\ConeofPCS X}=\Pcoh X$.

Given PCSs $X$ and $Y$ we define a PCS $\Limpl XY$ by
$\Web{\Limpl XY}=\Web X\times\Web Y$ and $t\in\Pcohp{\Limpl XY}$ if
for all $u\in\Pcoh X$, one has $\Matapp tu\in\Pcoh Y$ where
$(\Matapp tu)_b=\sum_{a\in\Web X}t_{a,b}u_a$ (matrix application). The
proof that this is indeed a PCS, as well as the proof of most of the
next results can be found in~\cite{DanosEhrhard08}. Such matrices can
be composed: given $s\in\Pcohp{\Limpl XY}$ and
$t\in\Pcohp{\Limpl YZ}$, $\Matapp ts\in\Pcohp{\Limpl XZ}$ is defined
as an ordinary composition of (usually infinite-dimensional) matrices
$(\Matapp ts)_{a,c}=\sum_{b\in\Web Y}s_{a,b}t_{b,c}$. In that way we
define the category $\PCOH$ whose objects are the PCSs and
$\PCOH(X,Y)=\Pcohp{\Limpl XY}$ ($\Id\in\PCOH(X,X)$ is given by
$\Id_{a,b}=\Kronecker ab$). This category is symmetric monoidal
closed, and actually *-autonomous, with
$\Tens XY=\Orthp{\Limpl X{\Orth Y}}$ which satisfies
$\Web{\Tensp XY}=\Web X\times\Web Y$ and
$\Pcohp{\Tens XY}=\Biorth{\Eset{\Tens uv\St u\in\Pcoh X\text{ and
    }v\in\Pcoh Y}}$. It is also cartesian with product of the family
(at most countable) $(X_i)_{i\in I}$ given by $\IWith_{i\in I}X_i=X$
where $\Web X=\cup_{i\in I}\Eset i\times\Web{X_i}$ and $v\in\Pcoh X$
if $v\in\Realpto{\Web X}$ satisfies
$\forall i\in I\ (v_{i,a})_{a\in\Web{X_i}}\in\Pcoh{X_i}$ and
projection $\Proj i\in\PCOH(X,X_i)$ given by
$(\Proj i)_{(j,b),a}=\Kronecker ji\Kronecker ba$.

\paragraph{PCSs as cones.}
There is a fully faithful functor $\ConeofPCS:\PCOH\to\SCCLin$
which maps a PCS $X$ to $\ConeofPCS X$ and a matrix $t\in\PCOH(X,Y)$
to the map $\ConeofPCS(t):\ConeofPCS X\to\ConeofPCS Y$ defined by
$\ConeofPCS(t)(x)=\Matapp tx$.  We use $\Clinfty$ for the full
subcategory of $\PCOH$ whose objects are the PCSs $X$ such that
$\Pcoh X=\Eset{u\in\Realpto{\Web X}\St u_a\leq 1}$.  This category
contains in particular the objects $\One$ (with
$\Web\One=\Eset\Onelem$), $\Orth\Snat$ and is closed under
$\IWith$. Notice that $\PCOH$ (and hence $\Clinfty$) is essentially
small since we only consider PCSs with at most countable webs so we
can assume that their webs are all subsets of $\Nat$ (in the sequel we
consider $\Clinfty$ as small).  We use $\ConeofLPCS$ for the inclusion
functor $\Clinfty\to\SCCLin$ (it is simply the restriction of
$\ConeofPCS$, so quite often we will drop the subscript $\infty$).

\begin{lemma}\label{lemma:pcoh-closed-charact}
  Let $I$ be an at most countable set and let $\cU\subseteq\Realpto I$
  be such that $\forall a\in I\ 0<\sup\Eset{u_a\St
    u\in\cU}<\infty$. Then $(I,\cU)$ is a PCS iff $\cU$ is convex,
  downwards closed and closed under lubs of monotone sequences.
\end{lemma}
This characterization was already stated and sketchily proven
in~\cite{Girard04a}, we provide a proof in the Appendix section
because it will be quite useful in the proof of the next Lemma.

\begin{lemma}\label{lemma:pcoh-as-closures}
  Let $I$ be an at most countable set and let $\cU\subseteq\Realpto I$
  be such that $\forall a\in I\ 0<\sup\Eset{u_a\St
    u\in\cU}<\infty$. Let $P$ be a cone and let $h:I\to P$ be such
  that $\forall u\in\cU\ \sum_{a\in I}u_ah(a)\in\Cuball P$. Then
  $\forall u\in\Biorth\cU\ \sum_{a\in I}u_ah(a)\in\Cuball P$ and
  $\bar h:u\mapsto\sum_{a\in I}u_ah(a)$ belongs to
  $\SCCLin(\ConeofPCS(I,\Biorth\cU),P)$.
\end{lemma}

Notice that \emph{it is not true} that $\Biorth\cU$ is the set of all
(at most countable) convex combinations $\sum_{j\in J}\alpha_j u(j)$
for $u(j)\in\cU$, simply because the set of these convex combinations
is not downwards closed in general.

\begin{example}\label{ex:tensor-diff}
  To illustrate this fact, take $I=\Eset{1,2}\times\Eset{1,2}$ and
  $\cU=\Eset{\Tens uv\St u,v\in\Pcohp{\With\One\One}}$ so that
  $(I,\Biorth\cU)=\Tens{(\With\One\One)}{(\With\One\One)}$. In the set
  $\cV$ of convex combinations of elements of $\cU$ we have for
  instance $\Base{1,1}=\Tens{\Base 1}{\Base 1}$,
  $\Base{2,2}=\Tens{\Base 2}{\Base 2}$,
  $\Base{1,1}+\Base{1,2}+\Base{2,1}+\Base{2,2}=\Tens{(\Base 1+\Base
    2)}{(\Base 1+\Base 2)}$, but we do not have
  $\Base{1,2}+\Base{2,1}$ (which cannot be obtained as a convex
  combination of $\Base{1,2}$ and $\Base{2,1}$). Notice that this
  latter element can be obtained as an iterated difference of convex
  combinations: $\Base{1,2}+\Base{2,1}=(\Tens{(\Base 1+\Base
    2)}{(\Base 1+\Base 2)}-\Tens{\Base 1}{\Base 1})-\Tens{\Base 2}{\Base 2}$.
\end{example}

\subsection{Density of probabilistic coherence spaces}
We prove that the functor $\ConeofLPCS$ is dense\footnote{This is
  already true if we replace $\ConeofLPCS$ with the full subcategory
  which has $\Orth\Snat$ as single object. Our formulation is
  motivated by Lemma~\ref{lemma:tensor-Clinfty}.}, in the sense
explained in Section~\ref{sec:gen-dense-functors}. Let
$P\in\Obj{\SCCLin}$, the objects of the category
$\Slice{\ConeofLPCS}{P}$ are the pairs $(X,f)$ where
$X\in\Obj\Clinfty$ and $f\in\SCCLin(\ConeofPCS X,P)$. And
$t\in\Slicep{\ConeofLPCS}{P}((X,f),(Y,g))$ means that $t\in\PCOH(X,Y)$
and $g\Compl\ConeofPCS(t)=f$. Then $\Slicepr P$ is the first
projection functor $\Slice{\ConeofPCS}{P}\to\SCCLin$ mapping $X$ to
$\ConeofPCS(X)$ and $t$ to $\ConeofPCS(t)$. And
$\Slicec P:\Slicepr P\Tonatural P$ is the cocone $(X,f)\mapsto f$.

Given $x\in P$ where $P\in\Obj\SCCLin$, we use $\Funofpt x$ for the
element of the cone $\Limpl{\ConeofPCS\One}P$ defined by
$\Funofpt x(\lambda)=\lambda x$ (so that
$\Norm{\Funofpt x}_{\Limpl{\ConeofPCS\One}P}=\Norm x$).

\begin{theorem}\label{th:Clinfty-dense-in-cones}
  The functor $\ConeofLPCS:\Clinfty\to\SCCLin$ is dense, that is, the
  cocone $\Slicec P$ is colimiting, for any object $P$ of $\SCCLin$.
\end{theorem}
\Beginproof
Let $\delta:\Slicepr P\Tonatural Q$ be another inductive cone. This
means that for each $X\in\Obj\Clinfty$ and each
$f\in\SCCLin(\ConeofLPCS X,P)$ we are given a
$\delta{(X,f)}\in\SCCLin(\ConeofLPCS X,Q)$ such that for any
$t\in\Clinfty(X,Y)$ we have the following implication of triangle
commutations:
\begin{center}
  \begin{tikzpicture}[->, >=stealth]
    \node (1) {$\ConeofPCSp X$};
    \node (mid) [right of = 1, node distance=10mm] {};
    \node (2) [right of=mid, node distance=10mm] {$\ConeofPCSp Y$};
    \node (3) [below of=mid, node distance=8mm] {$P$};
    \node (c) [right of=2, node distance=8mm] {};
    \node (i) [below of=c, node distance=4mm] {$\Implies$};
    \node (a1) [right of = 2, node distance=18mm] {$\ConeofPCSp X$};
    \node (amid) [right of = a1, node distance=10mm] {};
    \node (a2) [right of=amid, node distance=10mm] {$\ConeofPCSp Y$};
    \node (a3) [below of=amid, node distance=8mm] {$Q$}; 
    \tikzstyle{every node}=[midway,auto,font=\scriptsize]
    \draw (1) -- node {$\ConeofPCS(t)$} (2);
    \draw (1) -- node [swap] {$f$} (3);
    \draw (2) -- node {$g$} (3);
    \draw (a1) -- node {$\ConeofPCS(t)$} (a2);
    \draw (a1) -- node [swap] {$\delta{(X,f)}$} (a3);
    \draw (a2) -- node {$\delta{(Y,g)}$} (a3);
  \end{tikzpicture}
  \end{center}
In other words for all $t\in\Clinfty(X,Y)$
and $g\in\SCCLin(\ConeofPCS Y,P)$
\begin{align}\label{eq:cone-monoid-commut}
  \delta(X,g\Compl \ConeofPCS(t))=\delta(Y,g)\Compl \ConeofPCS(t)\,.
\end{align}

We first build a function $k:P\to Q$ so let $x\in P$. Assume first
that $\Norm x\leq 1$. Then $\Funofpt
x\in\SCCLin(\ConeofPCS\One,P)$. We set
\begin{align*}
  k(x)=\delta(\One,\Funofpt x)(\Base\Onelem)
\end{align*}
(remember that $\Onelem$ is the sole element of $\Web\One$) so that
$\Norm{k(x)}_Q\leq 1$ since
$\delta(\One,\Funofpt x)\in\SCCLin(\ConeofPCS\One,Q)$. Notice that if
$\lambda\in\Interval 01$ we have $\lambda\Id\in\Clinfty(\One,\One)$
and hence
\begin{align*}
  k(\lambda x)=\delta(\One,\Funofpt{\lambda x})(\Base\Onelem)
  &=\delta(\One,\Funofpt x\Compl(\lambda\Id))(\Base\Onelem)\\
  &=\delta(\One,\Funofpt x)(\lambda\Base\Onelem)=\lambda k(x)
\end{align*}
by~\Eqref{eq:cone-monoid-commut} and linearity of
$\delta(\One,\Funofpt x)$. Notice that we should have written
$\ConeofPCS(\lambda\Id)$ instead of $\lambda\Id$ in the formulas
above, we will systematically keep the $\ConeofPCS$
implicit\footnote{That is, consider morphisms of $\PCOH$ as morphisms
  of $\SCCLin$.} in this context to increase readability.

Therefore, given $x\in P$ we can set $k(x)=\Inv\lambda k(\lambda x)$
where $\lambda\in\Intervaloc 01$ is such that $\lambda\Norm x\leq 1$;
by the property we have just proven, this definition of $k(x)$ does
not depend on the choice of $\lambda$. Notice that
$\forall x\in P\ \Norm{k(x)}\leq\Norm x$ (since this holds when
$\Norm x=1$) and that $k(\lambda x)=\lambda k(x)$ holds for all
$x\in P$ and $\lambda\in\Realp$, that is, $k$ is homogeneous.

Now we prove that the function $k$ is linear. Let $x_1,x_2\in P$, we
must prove that $k(x_1+x_2)=k(x_1)+k(x_2)$. Since $k$ is homogeneous
we can assume that $\Norm{x_1}+\Norm{x_2}\leq 1$. Let
$a:\ConeofPCS(\With\One\One)\to P$ be defined by $a(u)=u_1x_1+u_2x_2$
(where $1,2$ are the elements of $\Web{\With\One\One})$.  This map is
linear, continuous (by continuity of scalar multiplication and
addition in $P$) and satisfies $\Norm a\leq 1$ by our assumption on
the $x_i$'s, hence $a\in\SCCLin(\ConeofPCSp{\With\One\One},P)$.

For $i=1,2$ we have
$k(x_i)=\delta(\One,\Funofpt{x_i})(\Base
\Onelem)=\delta(\One,a\Compl\Funofpt{\Base
  i})(\Base\Onelem)=\delta(\With\One\One,a)(\Base i)$
by~\Eqref{eq:cone-monoid-commut} (and the fact that
$\Funofpt{x}(\Base\Onelem)=x$). Hence
$k(x_1)+k(x_2)=\delta(\With\One\One,a)(\Base 1+\Base 2)$ by linearity
of $\delta(\With\One\One,a)$. Applying
again~\Eqref{eq:cone-monoid-commut}, as well as the definition of $a$,
we get
$k(x_1+x_2)=\delta(\One,\Funofpt{x_1+x_2})(\Base\Onelem)
=\delta(\One,a\Compl\Funofpt{\Base 1+\Base
  2})(\Base\Onelem)=\delta(\With\One\One,a)(\Base 1+\Base 2)$ which
proves our contention.

Next we prove that $k$ is continuous, so let $x(0)\leq x(1)\leq\cdots$
be a non-decreasing sequence in $\Cuball P$ and let $x\in\Cuball P$ be
its lub. For each $n\in\Nat$ we set $y(n)=x(n)-x(n-1)$ (we set
$x(-1)=0$ for convenience).

Let $u\in\ConeofPCSp{\Orth\Snat}$: this means that
$u\in\Realpto\Nat$ and $\sup_{n\in\Nat}u_n<\infty$. Let
$\lambda\in\Realp$ be such that $\forall n\in\Nat\
u_n\leq\lambda$. For each $N\in\Nat$ we have in $P$
\begin{align*}
  \sum_{n=0}^N u_ny(n)\leq\sum_{n=0}^N\lambda y(n)=\lambda x(N)\leq\lambda x
\end{align*}
and hence the non-decreasing sequence
$(\sum_{n=0}^N u_ny(n))_{N\in\Nat}$ has a lub in $P$ which is
$\sum_{n=0}^\infty u_ny(n)$, see Section~\ref{sec:cones-lin-cont}. So
we can define a function
\begin{align*}
  s:\ConeofPCSp{\Orth\Snat} \to P,\quad u \mapsto\sum_{n=0}^\infty u_ny(n)\,.
\end{align*}
Notice that $\forall u\in\Pcohp{\Orth\Snat}\ \Norm{s(u)}\leq\Norm x\leq 1$
since $s(u)\leq x$.

This map $s$ is linear by continuity of the algebraic operations of
$P$. We prove that it is continuous so let $(u(q))_{q\in\Nat}$ be a
non-decreasing sequence in $\Pcohp{\Orth\Snat}$ and let
$u\in\Pcohp{\Orth\Snat}$ be its lub (that is
$u_n=\sup_{q\in\Nat}u(q)_n$ for each $n\in\Nat$). We already know that
$\sup_{q\in\Nat}s(u(q))\leq s(u)$ by linearity of $s$ (which implies
monotonicity) so let us prove that
$s(u)\leq\sup_{q\in\Nat}s(u(q))$. This results from the fact that for
any $N\in\Nat$ we have
\begin{align*}
  \sum_{n=0}^Nu_ny(n)=\sup_{q\in\Nat}\sum_{n=0}^Nu(q)_ny(n)
  \leq\sup_{q\in\Nat}s(u(q))
\end{align*}
where the first equation results from the continuity of the algebraic
operations of $P$.

We have
$\delta(\One,\Funofpt{y(n)})(\Base\Onelem)
=\delta(\One,s\Compl\Funofpt{\Base n})(\Base\Onelem)
=\delta(\Orth\Snat,s)(\Base n)$ by~\Eqref{eq:cone-monoid-commut} (we
use also the observation that, setting
$u=\lambda\Base\Onelem\in\ConeofPCS\One$, one has
$(s\Compl\Funofpt{\Base n})(u)=s(\lambda\Base
n)=\lambda y(n)=\Funofpt{y(n)}(u)$, by definition of $s$).  Let
$e(N)=\sum_{n=0}^N\Base n\in\Pcohp{\Orth\Snat}$ so that $s(e(N))=x(N)$. We
have
\begin{align*}
  k(x(N))
  &= k(\sum_{n=0}^Ny(n))
  = \sum_{n=0}^Nk(y(n))\quad\text{by linearity of }k\\
  &= \sum_{n=0}^N \delta(\Orth\Snat,s)(\Base n)
    \quad\text{what we have just proven}\\
  &= \delta(\Orth\Snat,s)(e(N))
    \quad\text{linearity of }\delta(\Orth\Snat,s)
\end{align*}
and since $\delta(\Orth\Snat,s)$ is continuous we have
$\sup_{N\in\Nat}k(x(N))=\delta(\Orth\Snat,s)(e)$ where
$e=\sum_{n\in\Nat}\Base n$ (that is $e_n=1$ for all $n\in\Nat$).

Next
$k(x) =\delta(\One,\Funofpt x)(\Base\Onelem)
=\delta(\Orth\Snat,s\Compl\Funofpt e)(\Base\Onelem)
=\delta(\Orth\Snat,s)(e)$ by~\Eqref{eq:cone-monoid-commut} (we use
also the observation that
$(s\Compl\Funofpt{e})(u)=s(\lambda e)=\lambda x=\Funofpt{x}(u)$ where
$u=\lambda\Base\Onelem$, by definition of $s$) which proves that
$k(x)=\sup_{N\in\Nat}k(x(N))$ and hence that $k$ is continuous, so
$k\in\SCCLin(P,Q)$.

Now we prove that $k$ is a morphism of inductive cones
$\Slicepr P\Tonatural\delta$, that is, for any
$X\in\Obj{\Clinfty}$ and 
$f\in\SCCLin(\ConeofPCS X,P)$, the following triangle commutes:
\begin{center}
  \begin{tikzpicture}[->, >=stealth]
    \node (1) {$\ConeofPCS X$};
    \node (mid) [below of = 1, node distance=8mm] {};
    \node (2) [left of=mid, node distance=10mm] {$P$};
    \node (3) [right of=mid, node distance=10mm] {$Q$};
    \tikzstyle{every node}=[midway,auto,font=\scriptsize]
    \draw (1) -- node [swap] {$f$} (2);
    \draw (1) -- node {$\delta(X,f)$} (3);
    \draw (2) -- node {$k$} (3);
  \end{tikzpicture}  
\end{center}
Let $u\in\ConeofPCSp{X}$, we have
\begin{align*}
k(f(u))
=\delta(\One,\Funofpt{f(u)})(\Base\Onelem)
=\delta(\One,f\Compl\Funofpt u)(\Base\Onelem)
=\delta(X,f)(u)
\end{align*}
 by~\Eqref{eq:cone-monoid-commut} (we use also the
observation that
$(f\Compl\Funofpt{u})(v)=f(\lambda u)=\lambda f(u)=\Funofpt{f(u)}(v)$,
where $v=\lambda\Base\Onelem$, by linearity of $f$).

We end the proof that $\Slicec P$ is a colimiting cocone by observing that
$k$ is unique with these properties since its very definition is just
a particular case of the commutation expressing that $k$ is a morphism
of inductive cones (for $f=\Funofpt x$ with $x\in P$).
\Endproof

\subsection{Completeness of the category of cones}

\begin{theorem}\label{th:cones-lin-complete}
  The category $\SCCLin$ is complete, well-powered and
  admits $\Realp$ as co-generating object.
\end{theorem}
\Beginproof
First let $(P_i)_{i\in I}$ be a family of cones (where $I$ is any
set). We already have defined a cone $P=\prod_{i\in I}P_i$ as the set
of all families $\Vect x=(x_i)_{i\in I}$ such that $x_i\in P_i$ and
$(\Norm{x_i}_{P_i})_{i\in I}$ is bounded.

Equipped with the algebraic laws defined pointwise, it is a
cancellative $\Realp$-semi-module. We endow it with the norm
$\Norm{\Vect x}=\sup_{i\in I}\Norm{x_i}_{P_i}$ which clearly satisfies
all required axioms. The cone order of $P$ coincides with the product
order which shows readily that $P$ is a complete cone.

Together with the usual projections $\Proj i:\SCCLin(P,P_i)$, this
cone $P$ is the cartesian product of the $P_i$'s as easily checked. As
usual, given $f_i\in\SCCLin(Q,P_i)$ for each $i\in I$ we use
$\Tuple{f_i}_{i\in I}$ for the morphism $f\in\SCCLin(Q,P)$ such that
$f(y)=(f_i(y))_{i\in I}$ which is well defined by our definition of
$\SCCLin$ which requires\footnote{Without this condition, the category
  $\SCCLin$ has only finite products \emph{a priori}.} that all linear
morphisms are bounded by $1$. To finish we check that $P$ is
separated, so let $\Vect x,\Vect y\in P$ be such that
$\Vect x\not=\Vect y$. Let $i\in I$ be such that $x_i\not=y_i$. Let
$x'\in\Dual{P_i}$ be such that
$\Eval{x_i}{x'}\not=\Eval{y_i}{x'}$. Then $x'\Compl\Proj i\in\Dual P$
separates $\Vect x$ from $\Vect y$.

Let $P$ and $Q$ be cones and let $f,g\in\SCCLin(P,Q)$. Let
$E=\Eset{x\in P\St f(x)=g(x)}$. By linearity of $f$ and $g$, this set
$E$ inherits the algebraic structure of cancellative
$\Realp$-semi-module from $P$. We use $e$ for the inclusion
$E\subseteq P$ which is a semi-module morphism. Given $x\in E$ we set
$\Norm x_E=\Norm x_P$, which clearly defines a norm on
$E$. Completeness of $E$ follows from the fact that $f$ and $g$ are
continuous: indeed let $(x(n))_{n\in\Nat}$ be a sequence of elements
of $E$ which is non-decreasing in $E$ and hence in $P$ and satisfies
$\forall n\in\Nat\ \Norm{x(n)}\leq 1$ (for the norm of $E$, that is,
for the norm of $P$). Let $x\in\Cuball P$ be the lub of the $x(n)$'s
in $P$, by continuity of $f$ and $g$ we have $f(x)=g(x)$ and hence
$x\in E$. We finish the proof by showing that $x$ is the lub of the
$x(n)$'s in $E$, so let $y\in E$ be such that $x(n)\leq_E y$ for all
$n\in\Nat$. We have $x(n)\leq_P y$ and hence $x\leq_P y$ since $x$ is
the lub of the $x(n)$'s in $P$. By linearity of $f$ and $g$ we have
$f(y-x)=f(y)-f(x)=g(y)-g(x)=g(y-x)$ and hence $y-x\in E$ which shows
that $x\leq_E y$ as contended. The fact that $x\in\Cuball E$ results
obviously from the definition of the norm of $E$.

Next we prove that $E$ is separated. Let $x,y\in E$ be such that
$x\not=y$. By separateness of $P$ there is an $x'\in\Dual P$ such
that $\Eval x{x'}\not=\Eval y{x'}$. Let $y'$ be the restriction of
$x'$ to $E$, we have $y'\in\Dual E$ because all operations in $E$
(including the lubs) are defined as in $P$ and of course $y'$
separates $x$ from $y$.

Last we check that $(E,e)$ is the equalizer
of $f$ and $g$ in $\SCCLin$: let $h\in\SCCLin(H,P)$ be such that
$f\Compl h=g\Compl h$, this means exactly that
$\forall u\in H\ h(u)\in E$ so that we have a function $h_0:H\to E$
such that $h=e\Compl h_0$ (actually $h_0=h$ but it is safer to use
distinct names). The linearity and continuity of $h_0$ results from
the fact that the operations of $E$ are defined as in $P$ (including
lubs). Last $h_0(\Cuball H)\subset E\cap\Cuball P=\Cuball E$. Uniqueness of $h_0$ with these properties is obvious.

This proves that the category $\SCCLin$ is complete. The fact that
$\Realp$ is a cogenerator results from the fact that all the objects
of $\SCCLin$ are separated.

We are left with proving that $\SCCLin$ is well-powered. This results
from the following simple observation.

Let $H$ be an object of $\SCCLin$ and $h\in\SCCLin(H,P)$ be a
mono. This implies that $h$ is an injective function. Indeed, let
$u,v\in H$ with $u\not=v$. Wlog.~we can assume that
$\Norm u,\Norm v\leq 1$.  We have
$\Funofpt u,\Funofpt v\in\SCCLin(\Realp,H)$ and
$\Funofpt u(1)\not=\Funofpt v(1)$ hence $\Funofpt u\not=\Funofpt v$
and therefore $h\Compl\Funofpt u\not=h\Compl\Funofpt v$ from which it
follows by linearity of $h\Compl\Funofpt u$ and $h\Compl\Funofpt v$
that $h(u)=h(\Funofpt u(1))\not=h(\Funofpt v(1))=h(v)$.

Let $H_1=h(H)$ (so that $h$ is a bijection between $H$ and $H_1$) and
equip $H_1$ with the addition and scalar multiplication of $P$ so that
$h$ becomes an isomorphism of $\Realp$-semi-module from $H$ to $H_1$
(by linearity of $h$). We endow $H_1$ with the norm defined by
$\Norm x_{H_1}=\Norm{\Funinv h(x)}_H$. The cone $H_1$ defined in that
way is isomorphic to $H$ in our category $\SCCLin$. Let $\cS$ be the
category whose objects are the objects of $\SCCLin$ which, as sets,
are subsets of $P$ and morphisms are the monos of $\SCCLin$ (that is,
the morphisms which are injective functions), we have shown that there
is an equivalence between $\cS$ and the category of subobjects of $P$
(by the operation $(H,h)\mapsto H_1$ described above), and since $\cS$
is small (because the collection of all possible norms on a given
$\Realp$-semi-module is a set, and $\cS$ is locally small because
$\SCCLin$ is), this shows that $\SCCLin$ is well-powered.
\Endproof

\begin{theorem}\label{th:functor-has-adjoint}
  Any limit-preserving functor $F:\SCCLin\to\Cat C$, where the
  category $\Cat C$ is locally small, is a right adjoint.
\end{theorem}
\Beginproof
This is a direct application of the special adjoint functor theorem,
see~\cite{MacLane71} (Chap.~V, Sec.~8, Corollary).
\Endproof

\section{The tensor product of cones}
We use these categorical results to introduce the tensor product of
cones and prove its main properties.


\begin{lemma}\label{lemma:limpl-continuous}
  For any given object $P$ of $\SCCLin$, the functor
  $\Limpl P\Wcard:\SCCLin\to\SCCLin$ is continuous (that is, preserves
  all limits).
\end{lemma}

We are now in position of defining the tensor product of cones. For
the time being we use a notation different from the one we used for the
tensor product of PCSs.
\begin{theorem}\label{th:tens-as-adjoint}
  There is a unique functor $\ITensc:\SCCLin^2\to\SCCLin$ such that for
  each $P\in\Obj\SCCLin$, the functor $\Tensc\Wcard P$ is left adjoint
  to $\Limpl P\Wcard$ and that the bijection of the adjunction is
  natural in the three involved parameters.
\end{theorem}
\Beginproof
By Theorems~\ref{th:functor-has-adjoint} and~\ref{th:tens-as-adjoint},
for each $P\in\Obj\SCCLin$ the functor $\Limpl P\Wcard$ has a left
adjoint $\Tensc\Wcard P$. By the \emph{Adjunctions with a parameter}
theorem~\cite{MacLane71} (Chap.~IV, Sec.~7), this operation extends
uniquely to a functor $\SCCLin^2\to\SCCLin$ in such a way that the
bijection of the adjunction extends to a natural bijection
$\SCCLin(\Tensc{P_1}{P_2},Q)\Tonatural\SCCLin(P_1,\Limpl{P_2}Q)$ of
functors $\Op\SCCLin\times\Op\SCCLin\times\SCCLin\to\SET$.
\Endproof

\paragraph{Classification of bilinear maps.}
We refer to Section~\ref{sec:bilinear} for basic definitions on
bilinear maps.  We use $\Curlinc$ for the natural bijection
$\SCCLin(\Tensc RP,Q)\Tonatural\SCCLin(R,\Limpl PQ)$. We set
\begin{align*}
  \Tenscbil_{P,Q}=\Bilunc(\Curlinc(\Id_{\Tensc PQ}))\in\CCBilin
  PQ{\Tensc PQ}
\end{align*}
and we use also the notation $\Tensc xy$ for
$\Tenscbil_{P,Q}(x,y)\in\Tensc PQ$ (for $x\in P$ and $y\in Q$).

\begin{theorem}\label{th:tensc-classifies-bilin}
  Let $P$, $Q$ and $R$ be objects of $\SCCLin$.  For any
  $f\in\CCBilin PQR$ there is exactly one
  $\Linofbilin f\in\SCCLin(\Tensc PQ,R)$ such that
  $\Linofbilin f\Compl\Tenscbil=f$.
\end{theorem}
\Beginproof
We set $\Linofbilin f=\Funinv\Curlin(\Funinv\Bilunc(f))$. We have
\begin{align*}
  \Linofbilin f\Compl\Tenscbil
  &=\Linofbilin f\Compl\Bilunc(\Curlinc(\Id))\\
  &=\Bilunc(\Limplp Q{\Linofbilin f}\Compl\Curlinc(\Id))
    \quad\text{by naturality of }\Bilunc\\
  &=\Bilunc(\Curlinc(\Linofbilin f))\quad\text{by naturality of }\Curlinc\\
  &=f\,.
\end{align*}
Now we prove uniqueness so let $h\in\SCCLin(\Tensc PQ,R)$ be such that
$h\Compl\Tenscbil=f$. By the same kind of computation we have
$\Bilunc(\Curlinc(h))=\Bilunc(\Limplp
Qh\Compl\Curlinc(\Id))=h\Compl\Bilunc(\Curlinc(\Id))=h\Compl\Tenscbil=f$
from which it follows that $h=\Linofbilin f$.
\Endproof
This important universal property is however not sufficient for
proving that $\ITensc$ defines a monoidal structure on $\SCCLin$.  One
might solve this problem by showing that the natural
bijection $\SCCLin(\Tensc PQ,R)\Tonatural\SCCLin(P,\Limpl QR$ is
actually a natural isomorphism
$\Limplp{\Tensc PQ}{R}\Tonatural\Limplp P{\Limplp QR}$ of functors
$\Op\SCCLin\times\Op\SCCLin\times\SCCLin\to\SCCLin$. This almost
works, the only non trivial point seems to be the fact that the
inverse of this map has norm $\leq 1$ (we would probably need more
information about the elements of $\Cuballp{\Tensc PQ}$).

\paragraph{Action of $\ITensc$ on probabilistic coherence spaces.}
We use another method, based on the density of PCSs in cones that we
have proven; on the way we also learn that our new tensor product
coincides with the old one on PCSs.

\begin{theorem}
  There is a natural isomorphism
  \begin{align*}
    \Tensciso_{X,Y}:\ConeofPCS\Tensp XY\Tonatural\Tenscp{\ConeofPCS
    X}{\ConeofPCS Y}
  \end{align*}
  of functors $\PCOH^2\to\SCCLin$.
\end{theorem}
\Beginproof
Let
$\theta\in\CCBilin{\ConeofPCS X}{\ConeofPCS Y}{\ConeofPCS(\Tens XY)}$
be defined by $\theta(u,v)=\Tens uv$ (it is the bilinear
continuous map associated with the canonical morphism
$\Limpl X{\Limplp Y{\Tens XY}}$ in $\PCOH$). By
Theorem~\ref{th:tensc-classifies-bilin} we have an associated
$\Linofbilin\theta\in\SCCLin(\Tensc{\ConeofPCS X}{\ConeofPCS
  Y},\ConeofPCS(\Tens XY))$.  Now we define
$\rho\in \SCCLin(\ConeofPCS(\Tens XY),\Tensc{\ConeofPCS X}{\ConeofPCS
  Y})$.

Remember that
$\Pcohp{\Tens XY}=\Biorth{\Eset{\Tens uv\St u\in\Pcoh X\text{ and
    }v\in\Pcoh Y}}$ (\emph{warning}: $\Tens uv$ is the element of
$\Pcohp{\Tens XY}$ defined by $\Tensp uv_{a,b}=u_av_b$, not to be
confused, for the time being, with
$\Tensc uv\in\Tensc{\ConeofPCS X}{\ConeofPCS Y}$). Given $u\in\Pcoh X$
and $v\in\Pcoh Y$ we have
\begin{align*}
  \sum_{a\in\Web X,b\in\Web Y}\Tensp uv_{a,b}\Tenscbil(\Base a,\Base
  b)=\Tensc uv\in\Cuball{\Tenscp{\ConeofPCS X}{\ConeofPCS X}}
\end{align*}
by bilinearity and separate continuity of $\Tenscbil$.
By Lemma~\ref{lemma:pcoh-as-closures}
\begin{align*}
  \sum_{a\in\Web X,b\in\Web Y}w_{a,b}\Tenscbil(\Base a,\Base
  b)\in\Cuball{\Tenscp{\ConeofPCS X}{\ConeofPCS X}}  
\end{align*}
for all $w\in\Pcoh{\Tensp XY}$ and the map
\begin{align*}
  \rho:w\mapsto\sum_{a\in\Web X,b\in\Web Y}w_{a,b}\Tenscbil(\Base
  a,\Base b) 
\end{align*}
is linear and continuous
$\ConeofPCS\Tensp XY\to\Tensc{\ConeofPCS X}{\ConeofPCS Y}$ and
$\Norm\rho\leq 1$.

For $a\in\Web X$ and $b\in\Web Y$, we have
$\Linofbilin\theta(\rho(\Base{a,b}))=\Linofbilin\theta(\Tensc{\Base
  a}{\Base b})=\Tens{\Base a}{\Base b}=\Base{a,b}$ so that
$\Linofbilin\theta\Compl\rho=\Id$ by linearity and continuity. Next
for $u\in\Pcoh X$ and $v\in\Pcoh Y$ we have
$\rho\Compl\Linofbilin\theta\Compl\Tenscbil(u,v)
=\rho\Compl\theta(u,v)=\rho(\Tens
uv)=\Tensc uv=\Tenscbil(u,v)$ and hence
$\rho\Compl\Linofbilin\theta=\Id$ by the uniqueness part of the
universal property satisfied by $\Tenscbil$.
Naturality of $\Linofbilin\theta$ follows from its definition.
\Endproof

\paragraph{Cocontinuity of $\ITensc$.}
There is a natural transformation
\[
  \Symlimpl_{P_1,P_2,Q}\in
  \SCCLin(\Limpl{P_1}{\Limplp{P_2}{Q}},\Limpl{P_2}{\Limplp{P_1}{Q}})
\]
(of functors $\Op\SCCLin\times\Op\SCCLin\times\SCCLin\to\SCCLin$) from
which we derive a natural isomorphism
$\phi_{P_1,P_2,Q}:
\SCCLin(\Tensc{P_1}{P_2},Q)\Tonatural\SCCLin(\Tensc{P_2}{P_1},Q)$ by
Theorem~\ref{th:tens-as-adjoint} and by the fact that there is a
natural isomorphism $\SCCLin(P,Q)\Tonatural\SCCLin(\One,\Limpl PQ))$.
By Lemma~\ref{lemma:functor-yoneda-iso} we get a natural isomorphism
$\Symc_{P_1,P_2}\in\SCCLin(\Tensc{P_1}{P_2},\Tensc{P_2}{P_1})$.

\begin{theorem}\label{ref:tensc-separately-cocontinuous}
  The bifunctor $\ITensc:\SCCLin^2\to\SCCLin$ is separately cocontinuous.
\end{theorem}
\Beginproof
Being a left adjoint, the functor $\Tensc\Wcard P$ is cocontinuous. By
the existence of the natural isomorphism $\Symc$, it follows that
$\ITensc$ is cocontinuous separately in both parameters.
\Endproof

\subsection{Associativity isomorphisms of the tensor product}
We lift associativity of $\ITens$ on $\PCOH$ (more precisely
on the smaller category $\Clinfty$) to associativity of $\ITensc$ on
$\SCCLin$ by density.

\renewcommand\ConeofLPCS{\ConeofPCS}

\begin{lemma}\label{lemma:tensor-Clinfty}
  If $X$ and $Y$ are objects of $\Clinfty$ then $\Tens XY$ is also an
  object of $\Clinfty$.
\end{lemma}
\Beginproof
For a set $I$, let $1_I\in\Realpto I$ be defined by $(1_I)_i=1$ for
all $i\in I$. If $X$ is an object of $\Clinfty$ then
$1_{\Web X}\in\Pcoh X$ and hence
$1_{{\Web X}\times{\Web Y}}=\Tens{1_{\Web X}}{1_{\Web Y}}\in\Pcoh{\Tensp
  XY}$. If $w\in\Pcohp{\Tens XY}$ we must have $w_{a,b}\leq 1$ because
$\Base{(a,b)}\in\Pcoh{\Orthp{\Tens XY}}$ since
$\Eval{\Tens uv}{\Base{(a,b)}}\leq 1$ for all $u\in\Pcoh X$ and
$v\in\Pcoh Y$.
\Endproof

Given  $X_i\in\Obj\Clinfty$ for $i=1,2,3$, we define a natural
isomorphism
$\Assocli_{X_1,X_2,X_3}\in\SCCLin(\Tensc{\Tenscp{\ConeofPCS
    X_1}{\ConeofPCS X_2}}{\ConeofPCS X_3}, \Tensc{\ConeofPCS
  X_1}{\Tenscp{\ConeofPCS X_2}{\ConeofPCS X_3}})$ as the following
composition of natural isomorphisms
\begin{center}
  \begin{tikzpicture}[->, >=stealth]
    \node (1) {$\Tensc{\Tenscp{\ConeofPCS X_1}{\ConeofPCS X_2}}
      {\ConeofPCS X_3}$};
    \node (2) [ right of=1, node distance=54mm ]
      {$\Tensc{\ConeofPCS\Tensp{X_1}{X_2}}{\ConeofPCS X_3}$};
    \node (3) [ below of=2, node distance=10mm ]
      {$\ConeofPCS\Tensp{\Tensp{X_1}{X_2}}{X_3}$};
    \node (4) [ below of=1, node distance=10mm ]
      {$\ConeofPCS\Tensp{X_1}{\Tensp{X_2}{X_3}}$};
    \node (5) [ below of=4, node distance=10mm ]
      {$\Tensc{\ConeofPCS X_1}{\ConeofPCS\Tensp{X_2}{X_3}}$};
    \node (6) [ below of=3, node distance=10mm ]
      {$\Tensc{\ConeofPCS X_1}{\Tenscp{\ConeofPCS X_2}{\ConeofPCS X_3}}$};
    \tikzstyle{every node}=[midway,auto,font=\scriptsize]
    \draw (1) -- node {$\Tensc{\Tensciso_{X_1,X_2}}{\ConeofPCS X_3}$} (2);
    \draw (2) -- node {$\Tensciso_{\Tens{X_1}{X_2},X_3}$} (3);
    \draw (3) -- node [swap] {$\ConeofPCS\Assoc_{X_1,X_2,X_3}$} (4);
    \draw (4) -- node {$\Funinv{\Tensciso_{X_1,\Tens{X_2}{X_3}}}$} (5);
    \draw (5) --
      node {$\Tensc{\ConeofPCS X_1}{\Funinv{\Tensciso_{X_2,X_3}}}$} (6);
  \end{tikzpicture}  
  \end{center}

Now observe that both functors $T,T':\SCCLin^3\to\SCCLin$ defined
respectively by $T(P_1,P_2,P_3)=\Tensc{\Tenscp{P_1}{P_2}}{P_3}$ and
$T'(P_1,P_2,P_3)=\Tensc{P_1}{\Tenscp{P_2}{P_3}}$ (and similarly on
morphisms) are separately cocontinuous, because $\ITensc$ is
separately cocontinuous, see
Theorem~\ref{ref:tensc-separately-cocontinuous}. We have just
exhibited a natural isomorphism
$\Assocli:T\Compl\ConeofLPCS^3\Tonatural T'\Compl\ConeofLPCS^3$.
Since the functor $\ConeofPCS:\Clinfty\to\SCCLin$ is dense by
Theorem~\ref{th:Clinfty-dense-in-cones}, we can apply
Theorem~\ref{th:dense-concont-tnat-extension} which shows that there
is exactly one natural isomorphism $\Assocc:T\Tonatural T'$ such that
$\Assocc\Compl\ConeofLPCS^3=\Assocli$. In other words, there are
uniquely defined natural isomorphisms
$\Assocc_{P_1,P_2,P_3}\in\SCCLin(\Tensc{\Tenscp{P_1}{P_2}}{P_3},
\Tensc{P_1}{\Tenscp{P_2}{P_3}})$ such that, for all objects
$X_1,X_2,X_3\in\Obj\Clinfty$, one has
$\Assocc_{\ConeofPCS X_1,\ConeofPCS X_2,\ConeofPCS
  X_3}=\Assocli_{X_1,X_2,X_3}$.

Using the naturalities of $\Tensciso$, $\Assocc$ and $\Assoc$, and the
fact that $\Assoc$ satisfies MacLane Pentagon diagram in $\Clinfty$,
diagram chasing (Section~\ref{sec:pentagon}) shows that $\Assocc$
makes the following diagram commutative for any objects $X_i$
($i=1,2,3,4$) of $\Clinfty$
\begin{center}
  {\small
  \begin{tikzpicture}[->, >=stealth]
    \node (1) {$\Tensc{\Tenscp{\Tenscp{\ConeofPCS X_1}
          {\ConeofPCS X_2}}{\ConeofPCS X_3}}{\ConeofPCS X_4}$};
    \node (2) [ right of=1, node distance=46mm ]
    {$\Tensc{\Tenscp{\ConeofPCS X_1}{\ConeofPCS X_2}}
      {\Tenscp{\ConeofPCS X_3}{\ConeofPCS X_4}}$};
    \node (3) [ below of=1, node distance=20mm ]
    {$\Tensc{\Tenscp{\ConeofPCS X_1}{\Tenscp{\ConeofPCS X_2}
          {\ConeofPCS X_3}}}{\ConeofPCS X_4}$};
    \node (4) [ below of=2, node distance=20mm ]
    {$\Tensc{\ConeofPCS X_1}{\Tenscp{\Tenscp{\ConeofPCS X_2}
          {\ConeofPCS X_3}}{\ConeofPCS X_4}}$};
    \node (2bis) [ right of=2, node distance=30mm ] {};
    \node (5) [ below of=2, node distance=10mm ]
    {$\Tensc{\ConeofPCS X_1}{\Tenscp{\ConeofPCS X_2}
        {\Tenscp{\ConeofPCS X_3}{\ConeofPCS X_4}}}$};
    \tikzstyle{every node}=[midway,auto,font=\scriptsize]
    \draw (1) -- (2);
    \draw (1) -- (3);
    \draw (3) -- (4);
    \draw (4) -- (5);
    \draw (2) -- (5);
  \end{tikzpicture}
  }
  \end{center}
where the various morphisms are defined using $\Assocc$. This means
that the natural isomorphisms
\begin{align*}
  \psi^1_{\Vect P},\psi^2_{\Vect
  P}:\Tensc{\Tenscp{\Tenscp{P_1}{P_2}}{P_3}}{P_4}
  \Tonatural\Tensc{P_1}{\Tenscp{P_2}{\Tenscp{P_3}{P_4}}}
\end{align*}
defined by
\begin{align*}
  \psi^1_{\Vect P}&=\Assocc_{P_1,P_2,\Tensc{P_3}{P_4}}
                    \Compl\Assocc_{\Tensc{P_1}{P_2},P_3,P_4}\\
  \psi^2_{\Vect P}&=\Tenscp{P_1}{\Assocc_{P_2,P_3,P_4}}
                    \Compl \Assocc_{P_1,\Tensc{P_1}{P_2},P_3}
                    \Compl \Tenscp{\Assocc_{P_1,P_2,P_3}}{P_4}
\end{align*}
satisfy $\psi_1\Compl\ConeofLPCS^4=\psi_2\Compl\ConeofLPCS^4$ and
hence by the uniqueness statement of
Theorem~\ref{th:dense-concont-tnat-extension} we must have
$\psi^1=\psi^2$, that is, $\Assocc$ itself satisfies MacLane Pentagon
diagram. One deals similarly with the other coherence diagrams of
symmetric monoidal category (remember that we have defined a symmetry
natural isomorphism $\Symc$ in the proof of
Theorem~\ref{ref:tensc-separately-cocontinuous}, the other natural
isos $\Leftuc_P:\Tensc\One P\Tonatural P$ and
$\Rightuc_P:\Tensc P\One\Tonatural P$ are easy to define too).

We can summarize as follows what we have proven so far.

\begin{theorem}\label{th:cones-SMCC}
  The category $\SCCLin$ equipped with the tensor product $\ITensc$,
  the unit $\One$, the natural isos $\Leftuc$, $\Rightuc$, $\Assocc$
  and $\Symc$ is a symmetric monoidal category. It is closed, with
  object of morphisms from $P$ to $Q$ the cone $\Limpl PQ$ and
  evaluation $\Evlin\in\SCCLin(\Tensc{(\Limpl PQ)}{P},Q)$ induced by
  the bilinear and continuous map $(f,x)\mapsto f(x)$.
\end{theorem}

\section{The exponential}
Using again the special adjoint functor theorem, we equip $\SCCLin$
with a comonad $\Exclc\Wcard$ whose Kleisli category is (isomorphic
to) our category $\SCCStab$ of cones and stable functions. We start
with recalling the definition of the category.

Given $n\in\Nat$ we use $\Parte n$ (resp.~$\Parto n$) for the set of
all $I\subseteq\Eset{1,\dots,n}$ such that $n-\Card I$ is even
(resp.~odd).


Let $P$ and $Q$ be cones, in~\cite{EhrhardPaganiTasson18b} is
defined the notion of stable function $P\to Q$ and proven that cones
equipped with these functions form a cartesian closed category. Such a
function is defined only on $\Cuball P$,
\begin{itemize}
\item is bounded (that is $\Eset{\Norm{f(x)}\St x\in\Cuball P}$ is
  bounded),
\item totally monotone: for any $n\in\Nat$
  and any $\List x1n\in\Cuball P$ with
  $\sum_{i=1}^n x_i\in\Cuball P$, one has
  $\Appsumo f{\Vect x}\leq\Appsume f{\Vect x}$
  where $\Appsumo f{\Vect x}=\sum_{I\in\Parto n}f(\sum_{i\in I}x_i)$
  and $\Appsume f{\Vect x}=\sum_{I\in\Parte n}f(\sum_{i\in I}x_i)$ (notice
  that the conditions for $n=1,2$, namely $f(0)\leq f(x)$ and
  $f(x_1)+f(x_2)\leq f(x_1+x_2)+f(0)$, imply that $f$ is monotone),
\item and Scott-continuous (that is commutes with lubs of monotone
  sequences in $\Cuball P$).
\end{itemize}
Equipped with algebraic operations defined pointwise and with the norm
defined by $\Norm f=\sup_{x\in\Cuball P}\Norm{f(x)}$, the set of
stable functions is an object of $\SCCLin$ that we denote as
$\Simpl PQ$, separateness being proven as in the case of $\Limpl
PQ$. We use $\SCCStab$ for the category whose objects are those of
$\SCCLin$ and morphisms are the stable functions $f$ such that
$\Norm f\leq 1$.
\begin{theorem}[\cite{EhrhardPaganiTasson18b}]\label{th:cones-stab-CCC}
  The category $\SCCStab$ is cartesian closed with cartesian product
  defined as in $\SCCLin$, internal hom object $\Simpl PQ$ and
  evaluation map defined as in $\SET$.
\end{theorem}

Notice that $\SCCLin(P,Q)\subseteq\SCCStab(P,Q)$ since linearity
implies total monotonicity, this induces a ``forgetful'' faithful
functor $\Staboflin:\SCCLin\to\SCCStab$ which acts as the identity on
objects and morphisms. For the same reason we can consider
$\Simpl\Wcard\Wcard$ as a functor $\Op\SCCLin\times\SCCLin\to\SCCLin$
defined exactly in the same way as the functor $\Limpl\Wcard\Wcard$.

\begin{lemma}\label{lemma:stab-lin-exchange}
  With any $f\in\SCCLin(P,\Simpl QR)$ we can associate an element $g$
  of $\SCCStab(Q,\Limpl PR)$ defined by $g(y)(x)=f(x)(y)$. This
  correspondence is a natural bijection of functors
  $\Op\SCCLin\times\Op\SCCLin\times\SCCLin\to\SET$.
\end{lemma}
\Beginproof
Let $f\in\SCCLin(P,\Simpl QR)$. Let $y\in\Cuball Q$, the function
$f(\Wcard)(y):\ConeofPCS P\to\ConeofPCS R$ is linear and continuous
because the algebraic operations and lubs in $\Simpl QR$ are computed
pointwise. So it makes sense to define $g$ as in the statement of the
lemma, we must prove that this function is stable. First since $f$ is
linear and continuous, it is bounded so let $\lambda\in\Realp$ be such
that $\forall x\in\Cuball P\ \Norm{f(x)}_{\Simpl QR}\leq\lambda$. This
means that
$\forall x\in\Cuball P\,\forall y\in\Cuball Q\
\Norm{f(x)(y)}_Q\leq\lambda$. Therefore
$\forall y\in\Cuball Q\ \Norm{g(y)}_{\Limpl PR}\leq\lambda$. Next we
prove that $g$ is totally monotone so let $\List y1n\in\Cuball Q$ be
such that $\sum_{i=1}^ny_i\in\Cuball Q$. Let $x\in\Cuball P$, we have
\begin{align*}
  (\Appsumo g{\Vect y})(x)
  &=\Appsumo{(f(x))}{\Vect y}\quad\text{app.~is lin.~in the function}\\
  &\leq\Appsume{(f(x))}{\Vect y}\quad f(x)\text{ is stable} \\
  &=(\Appsume g{\Vect y})(x)
\end{align*}
and hence $\Appsumo g{\Vect y}\leq\Appsume g{\Vect y}$ since the
algebraic order of $\Limpl QR$ coincides with the pointwise
order. Continuity of $g$ follows similarly from that of each $f(x)$
and from the fact that lubs are computed pointwise in $\Limpl QR$.

Conversely let $g\in\SCCStab(Q,\Limpl PR)$. Let $x\in\ConeofPCS P$ and
let us check that the function $f(x)=g(\Wcard)(x)$ is stable. Let
$\lambda\in\Realp$ be such that
$\forall y\in\Cuball Q\ \Norm{g(y)}_{\Limpl PR}\leq\lambda$. Then we
have $\forall y\in\Cuball Q\ \Norm{g(y)(x)}_R\leq\lambda\Norm x_P$ and
this shows that $f(x)$ maps $\Cuball Q$ to a bounded subset of
$\ConeofPCS R$. Let $\List y1n\in\Cuball Q$ be such that
$\sum_{i=1}^ny_i\in\Cuball Q$, for the same reasons as above we have
$\Appsumo{(f(x))}{\Vect y}\leq\Appsume{(f(x))}{\Vect y}$ because
$\Appsumo g{\Vect y}\leq\Appsume g{\Vect y}$ by stability of
$g$. Therefore $f(x)$ is totally monotone. Continuity of $f(x)$
results from that of $g$ and from the fact that lubs are computed
pointwise in $\Limpl PR$. So $f(x)$ is well defined and belongs to
$\Simpl QR$. Now we prove that the function $f$ is linear. Let
$\List x1k\in P$ and $\List\alpha 1k\in\Realp$, we have
$\forall y\in\Cuball Q\
f(\sum_{j=1}^k\alpha_jx_j)(y)=\sum_{j=1}^k\alpha_jf(x_j)(y)$ by
linearity of each $g(y)$ and hence
$f(\sum_{j=1}^k\alpha_jx_j)=\sum_{j=1}^k\alpha_jf(x_j)$ because
algebraic operations are defined pointwise in $\Simpl QR$. Continuity
of $f$ holds for a similar reason.

These two operations are obviously natural and inverse of each other.
\Endproof

\begin{lemma}\label{lemma:func-staboflin-continuous}
  The functor $\Staboflin$ is continuous.
\end{lemma}

So by the special adjoint functor theorem $\Staboflin$ has a left
adjoint $\Linofstab:\SCCStab\to\SCCLin$. Let
$(\Exclc{},\Derc{},\Diggc{})$ be the associated comonad (in particular
$\Exclc{}=\Linofstab\Compl\Staboflin:\SCCLin\to\SCCLin$).


Let
$\Listadj_{P,Q}:\SCCLin(\Linofstab P,Q)\to\SCCStab(P,\Staboflin Q)$ be
the natural bijection associated with this adjunction. We have
$\Promfunc_P=\Listadj(\Id_{\Linofstab P})\in\SCCStab(P,\Exclc P)$
since $\Staboflin(\Linofstab P)=\Exclc P$; for any
$x\in\Cuball P$ we set $\Promc x=\Promfunc_P(x)\in\Cuballp{\Exclc
  P}$. This function $\Promfunc_P$ is the universal stable function:

\begin{lemma}\label{lemma:cones-prom-univ}
  For any $g\in\SCCStab(P,Q)$ there is exactly one function
  $\Linfactor g$ such that $g=\Linfactor g\Comp\Promfunc_P$, that is
  $\forall x\in\Cuball P\ g(x)=\Linfactor g(\Promc x)$. As a
  consequence, if $f_1,f_2\in\SCCLin(\Exclc P,Q)$ satisfy
  $\forall x\in\Cuball P\ \Matapp{f_1}{\Promc x}=\Matapp{f_2}{\Promc x}$
  then $f_1=f_2$.
\end{lemma}
\Beginproof
The first part is an immediate consequence of the adjunction, taking
$\Linfactor g=\Funinv{\Listadj_{P,Q}}(g)$ since
$g\in\SCCStab(P,\Staboflin Q)$. The second part a consequence of the
first for $g=f_1\Comp\Promfunc_P=f_2\Comp\Promfunc_P$.
\Endproof

\begin{lemma}
  Let $f\in\SCCLin(P,Q)$. Then $\Exclc f\in\SCCLin(\Exclc P,\Exclc Q)$
  is characterized by
  $\Matapp{\Exclc f}{\Promc x}=\Promc{(\Matapp fx)}$. Dereliction and
  digging are characterized by $\Matapp{\Derc{}}{\Promc x}=x$ and
  $\Matapp{\Diggc{}}{\Promc x}={\Prommc x}$.
\end{lemma}
These are direct consequences of the adjunction.  Given an unlabeled
binary tree $B$ with $n$ leaves and $\List P1n\in\Obj\SCCLin$, we use
$B(\List P1n)$ for the cone obtained by replacing each node of $B$
with the $\ITensc$ operator and the $i$th leaf with $\Exclc{P_i}$. For
instance if $B=\Pair{\Wcard}{\Pair\Wcard\Wcard}$ then
$B(P_1,P_2,P_3)=\Tensc{\Exclc{P_1}}{\Tenscp{\Exclc{P_2}}{\Exclc{P_3}}}$. We
define similarly $B(\List x1n)$ replacing the $i$th leaf with
$\Promc{x_i}$; in the example
$B(x_1,x_2,x_3)=\Tensc{\Promc{x_1}}{\Tenscp{\Promc{x_2}}{\Promc{x_3}}}$. The
next statement uses these notations.

\begin{lemma}\label{lemma:cones-prom-univ-multi}
  Let $f_1,f_2\in\SCCLin(B(\Vect P),Q)$ and assume that
  for any $x_1\in\Cuball{P_1}$,\dots,$x_n\in\Cuball{P_n}$, one has
  $f_1(B(\Vect x))=f_2(B(\Vect x))$ then $f_1=f_2$.
\end{lemma}
\Beginproof
By induction on $B$. If $B$ consists of one leaf this is just
Lemma~\ref{lemma:cones-prom-univ}. Assume $B=\Pair{B_1}{B_2}$ (with
$n=n_1+n_2$ and $B_i$ has $n_i$ leaves). Let $\Vect{P(i)}$ be a list
of cones of length $n_i$ (for $i=1,2$) and $\Vect P$ be the
concatenation of $\Vect{P(1)}$ and $\Vect{P(2)}$. We use similar
notations for elements of these cones. We have
$B(\Vect P)=\Tensc{B_1(\Vect{P(1)})}{B_2(\Vect{P(2)})}$ so that
$\Curlinc{f_j}\in\SCCLin(B_1(\Vect{P(1)}),\Limpl{B_2(\Vect{P(2)})}{Q})$
for $j=1,2$. Let $\Vect{x(1)}\in\Cuball{\Vect{P(1)}}$. For all
$\Vect{x(2)}\in\Cuball{\Vect{P(2)}}$ we have
\begin{align*}
  &(\Curlin{f_1})(B_1(\Vect{x(1)}))(B_2(\Vect{x(2)}))
  =f_1(B(\Vect x))\\
  &\hspace{1cm}=f_2(B(\Vect x))\quad\text{by the assumption on }f_1\text{ and }f_2\\
  &\hspace{1cm}=(\Curlin{f_2})(B_1(\Vect{x(1)}))(B_2(\Vect{x(2)}))
\end{align*}
and hence
$(\Curlin{f_1})(B_1(\Vect{x(1)}))=(\Curlin{f_2})(B_1(\Vect{x(1)}))$ by
inductive hypothesis applied to $B_2$. Next by inductive hypothesis
applied to $B_1$ we get $\Curlin{f_1}=\Curlin{f_2}$ and hence
$f_1=f_2$.
\Endproof


\begin{lemma}
  There is an iso $\Seelyzc\in\SCCLin(\One,\Exclc\Top)$ and a natural
  iso
  $\Seelytc_{P,Q}\in\SCCLin(\Tensc{\Exclc P}{\Exclc Q},\Exclc{\Withp
    PQ})$ such that $\Matapp{\Seelyzc}{1}=\Promc 0$ and
  $\Matapp\Seelytc{\Tenscp{\Promc x}{\Promc y}}=\Promc{(x,y)}$.
\end{lemma}
\Beginproof
We have a sequence of natural isomorphisms
\begin{align*}
  \SCCLin&(\Tensc{\Exclc P}{\Exclc Q},\Exclc{\Withp PQ})\\
  &\Tonatural \SCCLin(\Exclc P,\Limpl{\Exclc Q}{\Exclc{\Withp PQ}})
    \quad\text{by Theorem~\ref{th:cones-SMCC}}\\
  &\Tonatural \SCCStab(P,\Limpl{\Exclc Q}{\Exclc{\Withp PQ}})\quad\text{since }
    \Linofstab\Adj\Staboflin\\
  &\Tonatural \SCCLin(\Exclc Q,\Simpl{P}{\Exclc{\Withp PQ}})
    \quad\text{by Lemma~\ref{lemma:stab-lin-exchange}}\\
  &\Tonatural \SCCStab(Q,\Simpl{P}{\Exclc{\Withp PQ}})\\
  &\Tonatural \SCCStab(\With QP,{\Exclc{\Withp PQ}})\\
  &\Tonatural \SCCStab(\With PQ,{\Exclc{\Withp PQ}})
    \quad\text{by symmetry of }\IWith\\
  &\Tonatural \SCCLin(\Exclc{\Withp PQ},{\Exclc{\Withp PQ}})\ni\Id
\end{align*}
whence a natural
$\Seelytc_{P,Q}\in\SCCLin(\Tensc{\Exclc P}{\Exclc Q},\Exclc{\Withp
  PQ})$. This definition implies that
$\Matapp\Seelytc{\Tenscp{\Promc x}{\Promc y}}=\Promc{(x,y)}$. Next we
define
$f:\Cuball{\Withp PQ}\to\Cuball{\Tenscp{\Exclc P}{\Exclc Q}}$
by $f(x,y)=\Tensc{\Promc x}{\Promc y}$. This function is stable
because $\Promfunc$ is stable and $\ITensc$ is bilinear and
continuous. So we have
$\Linfactor f\in\SCCLin(\Exclc{\Withp PQ},\Tensc{\Exclc P}{\Exclc Q})$
which satisfies
$\Matapp{\Linfactor f}{\Promc{(x,y)}}=\Tensc{\Promc x}{\Promc y}$. By
Lemma~\ref {lemma:cones-prom-univ-multi} it follows that
$\Linfactor f$ is the inverse of $\Seelytc$.

Since $\Top=\Eset 0$ and $\One=\Realp$ we have
$g\in\SCCStab(\Top,\One)$ given by $g(0)=1$ and hence
$\Linfactor g\in\SCCLin(\Exclc\Top,\One)$ fully characterized by
$\Matapp{\Linfactor g}{\Promc 0}=1$. We define
$\Seelyzc\in\SCCLin(\One,\Exclc\Top)$ by
$\Seelyzc(\lambda)=\lambda\Promc 0$. Lemma~\ref
{lemma:cones-prom-univ-multi} shows that
$\Seelyzc\Compl\Linfactor g=\Id$ and $\Linfactor g\Compl\Seelyzc=\Id$
is straightforward.
\Endproof

\begin{theorem}
  Equipped with the above natural transformations
  $(\Derc{},\Diggc{},\Seelyzc,\Seelytc)$, the functor $\Exclc{}$ is a
  strong symmetric monoidal comonad from the symmetric monoidal
  category $(\SCCLin,\IWith)$ to the symmetric monoidal category
  $(\SCCLin,\ITensc)$ and $\SCCStab$ is equivalent to the Kleisli
  category of this comonad.
\end{theorem}
\Beginproof
This boils down to proving the commutation of a few diagrams
(see~\cite{Mellies09}) using the above characterizations of maps by
their action on tensors of elements of shape $\Promc x$.
\Endproof

\section{Measurability}
Let $\cX$ and $\cY$ be measurable spaces. A substochastic kernel
$\cX\Tokern\cY$ is a map $K:\cX\times\Sigma_\cY\to\Realp$ such that
for each $r\in\cX$, the map $K(r,\Wcard)$ is a subprobability measure
on $\cY$ and, for each $V\in\Sigma_\cY$, the map $K(\Wcard,V)$ is
measurable.  Such a kernel $K$ induces
$\Linofkern K\in\SCCLin(\Meas\cX,\Meas\cY)$ given by
$\Linofkern K(\mu)(V)=\int K(r,V)\mu(dr)$ from which $K$ can be
recovered since $K(r,V)=\Linofkern K(\Dirac r)(V)$ (where $\Dirac r$ is
the Dirac measure at $r$). It is not true however that any
$k\in\SCCLin(\Meas\cX,\Meas\cY)$ allows to define a kernel $K$ by
setting $K(r,V)=k(\Dirac r)(V)$ because there is no reason for this
function to be measurable in $r$. This is why the objects of $\SCCLin$
must be equipped with an additional measurability structure and the
linear and continuous morphisms must respect this structure. This set
of definitions is very close in spirit to quasi-Borel
spaces~\cite{VakarKammarStaton19}.

Let $\MSP$ be the category of measurable spaces and measurable
functions and let $\Mref:\MREFC\to\MSP$ be a functor from a cartesian
reference category $\MREFC$. We require $\Mref$ to preserve all finite
cartesian products. The choice of this reference functor depends on
the data-types of the language we want to interpret. If, as
in~\cite{EhrhardPaganiTasson18b}, the language has the real numbers as
ground type, one takes $\MREFC=\Nat$,
$\MREFC(n,m)=\MSP(\Realto n,\Realto m)$, $\Mref(n)=\Realto n$ and
$\Mref(h)=h$ for $h\in\MREFC(n,m)$. We use $\Rzero$ for the terminal
object of $\MREFC$ (in our example it is $0\in\Nat$) and,
with this example in mind, we use $+$ for the cartesian product in
$\MREFC$. Hence $\Mref(\Rzero)$ is the one-point measurable space and
$\Mref(p+q)=\Mref(p)\times\Mref(q)$. To simplify notations a little we
assume that, as in our motivating example, the functor $\Mref$ acts as
the identity on morphisms, that is
$\MREFC(p,q)=\MSP(\Mref(p),\Mref(q))$.

A \emph{measurable cone} is a pair $P=(\Mconec P,\Mconet P)$ where
$\Mconec P\in\Obj\SCCLin$ and $\Mconet P=(\Mconet P_p)_{p\in\MREFC}$
is a family of sets
$\Mconet P_p\subseteq(\Cdual{\Mconec P})^{\Mref(p)}$ whose element
satisfy\footnote{It is convenient to use $\Metabs\Wcard\Wcard$
  notation borrowed to the $\lambda$-calculus to write some of the
  involved functions.}:
%
if $l\in\Mconet P_p$ then
$\forall x\in\Mconec P\ \Metabs
r{l(r)(x)}\in\MSP(\Mref(p),\Realp)$. Moreover this family  is
closed under precomposition\footnote{This can be described in terms of
  presheaves of sets.} by morphisms in $\MREFC$: if $l\in\Mconet P_p$
then
$\forall h\in\MREFC(q,p)\ \Metabs s{\Metabs x{l(h(s))(x)}}\in\Mconet
P_q$. The $l\in\Mconet P_p$ are the \emph{measurability tests
  of arity $p$} of $P$.

A \emph{measurable path of arity $p$ of $P$} is a map
$\gamma:\Mref(p)\to\Cuballp{\Mconec P}$ such that,
for all $q\in\MREFC$ and all $m\in\Mconet P_q$ one has
$\Metabs{(r,s)}{m(s)(\gamma(r))}\in\MSP(\Mref(p+q),\Realp)$. We use
$\Mpath P_p$ for the set of these paths.  Notice that for any
$x\in\Cuball{\Mconec P}$ one has $\Metabs rx\in\Mpath P_p$ for any
$p$.  An $f\in\SCCLin(\Mconec P,\Mconec Q)$ is \emph{measurable} if
$\forall \gamma\in\Mpath P_p\ f\Comp\gamma\in\Mpath Q_p$. We use
$\SCCLinm$ for the category of measurable cones and measurable
continuous linear functions. Let $(P_i)_{i\in I}$ be a family of
measurable cones. Given $i\in I$ and $l\in\Mconet{P_i}_p$, we define
$\Inj i(l)$ as the element of
${\Cdual{(\prod_{j\in I}\Mconec{P_j})}}^{\Mref(p)}$ defined by
$\Inj i(l)(r)(\Vect x)=l(r)(x_i)$. We set\footnote{Slightly simpler
  definition than in~\cite{EhrhardPaganiTasson18b}, but the sets of
  measurable paths to $\prod_{i\in I}\Mconec{P_i}$ are the same. This
  also explains why we have dropped the first requirement on families
  of sets of measurability tests.}
$\Mconet{\prod_{i\in I}P_i}_p=\Eset{\Inj i(l)\St i\in I\text{ and
  }l\in\Mconet{P_i}_p}$ thus defining a measurable cone
$\prod_{i\in I}P_i$ which is easily seen to be, when equipped with the
ordinary projection maps, the cartesian product of the $P_i$'s in
$\SCCLinm$, so this category is cartesian\footnote{It would not be
  difficult to check that it is actually small-complete by showing
  that it has also binary equalizers}.

Let $\Mconec{\Limplm PQ}$ be the cone\footnote{It is easy to check
  that these functions equipped with the norm defined as in
  $\Limpl{\Mconec P}{\Mconec Q}$, is a cone. The only point which
  deserves a mention is the proof of completeness which uses in a
  crucial way the monotone convergence theorem; as mentioned
  in~\cite{EhrhardPaganiTasson18b} this explains why cones are
  complete only for bounded monotone sequences and not arbitrary
  directed families.}  of linear and continuous functions $P\to Q$
which are measurable in the sense that $\lambda f$ is measurable for some
$\lambda>0$. It is easy to check that one turns this cone into a
measurable cone $\Limplm PQ$ by equipping it with
$\Mconet{\Limplm PQ}_p=\Eset{\Lftest\gamma l\St\gamma\in\Mpath
  P_p\text{ and }l\in\Mconet Q_p}$ where
$\Lftest\gamma l=\Metabs r{\Metabs
  f{l(r)(f(\gamma(r)))}}\in{\Cdual{(\Mconec{\Limplm
      PQ})}}^{\Mref(p)}$.

Given two measurable cones $P$ and $Q$, we define $\Tensc PQ$ as the
measurable cone $(\Tensc{\Mconec P}{\Mconec Q},\Mconet{\Tensc PQ})$
where $m\in{\Cdual{\Tenscp{\Mconec P}{\Mconec Q}}}^{\Mref(k)}$ belongs
to $\Mconet{\Tensc PQ}_k$ if for all
$z\in\Tensc{\Mconec P}{\Mconec Q}$, one has
$\Metabs w{m(w)(z)}\in\MSP(\Mref(k),\Realp)$ and for all
$\gamma\in\Mpath P_p$ and $\delta\in\Mpath Q_q$
\begin{align*}
  \Metabs{(r,s,w)}{m(w)(\Tensc{\gamma(r)}{\delta(s)})}
  \in\MSP(\Mref(p+q+k),\Realp)\,.
\end{align*}
It is easily checked that
$(\Tensc{\Mconec P}{\Mconec Q},\Mconet{\Tensc PQ})$ is indeed a
measurable cone $\Tensc PQ$.

\begin{lemma}\label{lemma:tens-of-paths}
  Let $\gamma\in\Mpath P_q$ and $\delta\in\Mpath Q_q$. Then
  $\Metabs{(r,s)}{\Tensc{\gamma(r)}{\delta(s)}}\in\Mpath{\Tensc
    PQ}_{p+q}$, we use $\Tens\gamma\delta$ for this path.
\end{lemma}

\begin{lemma}\label{lemma:tens-measurable}
  Given measurable cones $P,Q,R$, the bijection
  $\Curlin:\SCCLin(\Tensc{\Mconec P}{\Mconec Q},\Mconec
  R)\to\SCCLin(\Mconec P,\Limpl{\Mconec Q}{\Mconec R})$ restricts to a
  bijection $\SCCLinm(\Tensc{P}{Q},R)\to\SCCLinm(P,\Limplm{Q}{R})$.
\end{lemma}

\begin{lemma}\label{lemma:bilinear-mesurable}
  Let $f\in\SCCLin(\Tensc{\Mconec P}{\Mconec Q},R)$. One has
  $f\in\SCCLinm(\Tensc PQ,R)$ iff for all $\gamma\in\Mpath P_p$ and
  $\delta\in\Mpath Q_q$, one has
  $f\Comp(\Tensc\gamma\delta)\in\Mpath{\Tensc PQ}_{p+q}$.
\end{lemma}
Immediate consequence of the above. It generalizes easily, replacing
$\Tensc PQ$ with any tensorial tree like
$\Tensc{P_1}{\Tenscp{P_2}{P_3}}$. It is then routine to prove the
following.

\begin{theorem}
  The functor $\ITensc$ restricts to a functor $\SCCLinm^2\to\SCCLinm$
  (still denoted $\ITensc$).  Equipped with $\ITensc$, the category
  $\SCCLin$ is symmetric monoidal closed.
\end{theorem}

\begin{example}
  Let $\cX$ be a measurable space (with $\Sigma$-algebra
  $\Sigma_\cX$). Given $p\in\MREFC$ and $U\in\Sigma_X$ we define
  $\Measev U\in(\Cdual{\Meas\cX})^{\Mref(p)}$ by
  $\Measev U(r)(\mu)=\mu(U)$ (for $r\in\Mref(p)$). The measurable cone
  $\Measm\cX$ is defined by $\Mconec{\Measm\cX}=\Meas\cX$ and
  $\Mconet{\Measm\cX}_p=\Eset{\Measev U\St U\in\Sigma_\cX}$. This
  means that $\Mpathu{\Measm\cX}_p$ is the set of all maps
  $\gamma:\Mref(p)\to\Meas\cX$ such that
  $\Eset{\gamma(r)(\cX)\St r\in\Mref(p)}\subseteq\Realp$ is bounded by
  $1$ and, for each $U\in\Sigma_\cX$, the map
  $\Metabs r{\gamma(r)(U)}$ is in $\MSP(\Mref(p),\Realp)$; in other
  words $\gamma:\Mref(p)\Tokern\cX$.  Let $K:\cX\Tokern\cY$, the
  associated map
  $\Linofkern K\in\SCCLin(\Mconec{\Measm\cX},\Mconec{\Measm\cY})$ is
  measurable because, given $\gamma\in\Mconet{\Measm\cX}_p$,
  $\Linofkern K\Comp\gamma$ is nothing but the usual composition of
  the substochastic kernels\footnote{We are implicitly using the Giry
    monad.} $K$ and $\gamma$. Conversely let
  $f\in\SCCLin(\Measm\cX,\Measm\cY)$ and assume that $\cX=\Mref(p)$
  for some $p\in\MREFC$. Then
  $\gamma=\Metabs r{\Dirac r}\in\Mpathu{\Measm\cX}_p$ (it is the
  identity kernel) and hence $K=f\Comp\gamma:\cX\Tokern\cY$ by
  measurability of $f$, which satisfies $\Linofkern K=f$. So if we
  take $\MREFC=\MSP$ and $\Mref=\Id$, the category of measurable
  spaces and substochastic kernels is a full subcategory of
  $\SCCLinm$. It seems clear that
  $\Measm{\cX\times\cY}=\Tensc{\Measm\cX}{\Measm\cY}$, with
  $\Tensc\mu\nu=\Tens\mu\nu$ (the usual tensor product of measures),
  this will be checked in further work.
\end{example}

\paragraph{The exponential.}
We only sketch this case which is quite similar to that of
$\ITensc$. As in~\cite{EhrhardPaganiTasson18b} we say that
$f\in\SCCStab(\Mconec P,\Mconec Q)$ is measurable if
$\forall\gamma\in\Mpathu P_q\ f\Comp\gamma\in\Mpath Q_p$. We use
$\SCCStabm$ for the category of measurable cones and measurable stable
functions, it is a CCC.

Let $P$ be a measurable cone. We define $\Mconet{\Exclc P}_q$ as the
set of all elements $l$ of ${\Cdual{(\Exclc{\Mconec P})}}^{\Mref(q)}$
such that for all
$z\in\Exclc{\Mconec P}$, $\Metabs s{l(s)(z)}\in\MSP(\Mref(q),\Realp)$
and for all
$\gamma\in\Mpath P_p$,
$\Metabs{(r,s))}{l(s)(\Promc{\gamma(r)})}\in\MSP(\Mref(p+q),\Realp)$. In
that way, as easily checked, we have defined a measurable cone
$\Exclc P$. If $\gamma\in\Mpath P_p$ then clearly
$\Promc\gamma=\Metabs r{\Promc{(\gamma(r))}}\in\Mpath{\Exclc P}_p$.

\begin{theorem}\label{lemma:exp-measurable}
  The bijection
  $\SCCLin(\Exclc{\Mconec P},\Mconec Q)\to\SCCStab(\Mconec P,\Mconec
  Q)$ restricts to a bijection
  $\SCCLinm(\Exclc P,Q)\to\SCCStabm(P,Q)$.
\end{theorem}

Hence, if $f\in\SCCLin(\Exclc{\Mconec P},\Mconec Q)$, then
$f\in\SCCLinm(\Exclc P,Q)$ iff for all $\gamma\in\Mpath P_p$, it holds
that $f\Comp\Promc\gamma\in\Mpath Q_p$. The required properties of
$\Exclc\Wcard$ and of its associated structures follow easily.

\section{Conclusion}
We need to understand better the internal structure of $\Tensc PQ$ and
$\Exclc P$ (without and with measurability structure), for instance as
suggested in the Introduction we conjecture that $\Tensc PQ$ is the
smallest subcone of $\Cdual{\SCCBilin PQ\Realp}$ which contains all
the operators $\Tens xy:f\mapsto f(x,y)$ on bilinear forms, and
similarly of $\Exclc P$.  We also conjecture that
$\Exclc{\ConeofPCS X}$ and $\ConeofPCS\Excl X$ are naturally
isomorphic (for PCSs $X$).

The framework of measurable complete positive cones seems now to be
quite a general and flexible one, allowing not only to interpret
probabilistic programming languages using continuous data types such
as the real line and also general recursive data-types (this feature
will be presented in a forthcoming paper), but also hosting naturally
differential operations on programs. For instance, given a stable
$f:\Cuball P\to\Realp$ and elements $x,u\in\Cuball P$ such that
$x+u\in\Cuball P$ we know thanks to~\cite{Crubille18} that the map
$\phi_u:[0,1]\to\Realp$, $\lambda\mapsto f(x+\lambda u)$ belongs to
$\PCOH(\Excl\One,\One)$ and hence has a derivative
$\phi'_u(0)\in\Realp$. The map $u\mapsto \phi'_u(0)$ is linear and
continuous $P_x\to\Realp$ (where $P_x$ is the ``local cone'' of $P$ at
$x$, that is the cone of all $u\in P$ such that
$x+\lambda u\in\Cuball P$ for some $\lambda>0$, equipped with a
suitable norm, the obvious generalization of a construction
of~\cite{Ehrhard19a} for PCSs) thus allowing to introduce a general
differential calculus for stable functions on cones with expected
applications in optimization as well as static analysis of
programs. Of course the linear constructs on cones of this paper will
be essential in these forthcoming developments.

Another interesting outcome of this work is the fact that PCSs are
dense in the category $\SCCLin$, a fact which might be quite useful for
transferring the full abstraction results obtained so far to
probabilistic programming languages handling continuous data-types.
The completeness of $\SCCLin$ might also be quite an useful feature
and an incentive for extending linear logic with dependent types; as
an illustration we exhibit a natural cone which arises as an equalizer
of two linear endomorphisms of a PCS.
\begin{example}
  Let $X$ be the least solution of the equation
  $X=\With\One{\Tensp\Snat X}$ in $\PCOH$ in the sense explained
  in~\cite{DanosEhrhard08}, it can be seen as a type of streams of
  integers. This PCS can be described simply: $\Web X$ is the set of
  \emph{finite} sequences of integers and $u\in\Realpto{\Web X}$ is in
  $\Pcoh X$ if $\sum_{a\in A}u_a\leq 1$ for all antichains
  $A\subseteq\Web X$ (that is, set of pairwise incomparable finite
  sequences). Then we have a morphism $s\in\PCOH(X,X)$ which is
  defined by $s_{a,b}=1$ if $a$ is of shape $b.n$ ($n\in\Nat$ added at
  the end of the sequence $b$) and $s_{a,b}=0$ otherwise. In other
  words $(\Matapp su)_b=\sum_{n\in\Nat}u_{b.n}$. It is not hard to see
  that the equalizer of $s$ and
  $\Id\in\PCOH(\ConeofPCS X,\ConeofPCS X)$ is isomorphic to $\Meas\cX$
  where $\cX$ is the Baire space (the Polish space $\Nat^\omega$)
  equipped with its Borel $\Sigma$-algebra: if $\Matapp su=u$ then $u$
  can be seen as the measure which maps the basic clopen set of all
  sequences $\in\Nat^\omega$ extending $a$ to $u_a$. It is even
  possible to check that the measurability structure introduced
  in~\cite{Crubille18} for general PCSs seen as cones induces a
  measurability structure on this equalizer such that its measurable
  paths are exactly the substochastic kernels to $\cX$. This example
  shows that equalizers of simply definable morphisms on recursively
  definable types can have quite an interesting structure.
\end{example}



\paragraph{Acknowledgments.}
  We thank R.~Crubillé, F.~Dahlqvist, P.-A.~Melliès, M.~Pagani and
  C.~Tasson for many illuminating discussions on these topics. This
  work has been partly funded by the ANR PRC project
  \emph{Probabilistic Program Semantics} (PPS).


\bibliography{../../newbiblio}

\section{Appendix}

\subsection{Proof of Lemma~\ref{lemma:functor-yoneda-iso}}

\Beginproof
We prove first naturality of $\eta$, so let $f\in\Cat C(C,C')$, we
have
\begin{align*}
  F(f)\Compl\eta_C
  &=(\Cat D(G(C),F(f))\Comp \psi_{C,F(C)})(\Id_{F(C)})\\
  &=(\psi_{C,F(C')}\Comp\Cat D(F(C),F(f)))(\Id_{F(C)})\\
  &=\psi_{C,F(C')}(F(f))\\
  &=(\psi_{C,F(C')}\Comp\Cat D(F(f),F(C')))(\Id_{F(C')})\\
  &=(\Cat D(G(f),F(C'))\Comp\psi_{C',F(C')})(\Id_{F(C')})\\
  &=\eta_{C'}\Compl G(f)
\end{align*}
by commutation of the diagrams
\begin{equation*}
  \begin{tikzpicture}[->, >=stealth]
    \node (1) {$\Cat D(F(C),F(C))$};
    \node (2) [right of=1, node distance=40mm] 
      {$\Cat D(G(C),F(C))$};
    \node (3) [below of=1, node distance=12mm]
      {$\Cat D(F(C),F(C'))$}; 
    \node (4) [below of=2, node distance=12mm]
    {$\Cat D(G(C),F(C'))$};
    \node (5) [below of=3, node distance=12mm]
    {$\Cat D(F(C'),F(C'))$};
    \node (6) [below of=4, node distance=12mm]
    {$\Cat D(G(C'),F(C'))$};
    \tikzstyle{every node}=[midway,auto,font=\scriptsize]
    \draw (1) -- node {$\psi_{C,F(C)}$} (2);
    \draw (1) -- node [swap]
    {$\Cat D(F(C),F(f))$} (3);
    \draw (2) -- node {$\Cat D(G(C),F(f))$} (4);
    \draw (3) -- node {$\psi_{C,F(C')}$} (4);
    \draw (5) -- node {$\Cat D(F(f),F(C'))$} (3);
    \draw (5) -- node {$\psi_{C',F(C')}$} (6);
    \draw (6) -- node [swap] {$\Cat D(G(f),F(C'))$} (4);
  \end{tikzpicture}  
\end{equation*}
and naturality of $\theta$ is similar. Next, by naturality of $\psi$
and definition of $\theta_C$ we have
\begin{align*}
  \theta_C\Compl\eta_C
  &= \theta_C\Compl\psi_{C,F(c)}(\Id_{F(C)})\\
  &= \Cat D(G(C),\theta_C)\Comp\psi_{C,F(C)})(\Id_{F(C)})\\
  &= (\psi_{C,G(C)}\Comp\Cat D(F(C),\theta_C))(\Id_{F(C)})\\
  &= \psi_{C,G(C)}(\theta_C)=\Id_{G(C)}
\end{align*}
The equation
$\eta_C\Compl\theta_C=\Id_{F(C)}$ is proven similarly.
\Endproof

\subsection{Proof of Lemma~\ref{lemma:functor-transpose-cocont}}
\Beginproof
Let $\Delta:J\to\Cat C$ be a diagram and $\gamma:\Delta\Tonatural c$ be a
colimiting cocone, we must prove that
$F'\gamma:F'\Delta\Tonatural F'(c)$ is a colimiting cocone in
$\Funcat{\Cat D}{\Cat E}$, so let $\delta:\Delta\Tonatural H$ be
another cocone based on $\Delta$ in $\Funcat{\Cat D}{\Cat E}$. For any
objects $j$ of $J$ and $d$ of $\Cat D$ we have that $(\delta_j)_d$
(which we simply denote as $\delta_{j,d}$) belongs to
$\Cat E(F(\Delta(j),d),H(d))$ and is natural in $j$ and $d$, that is,
for any $\phi\in J(j,j')$ and $g\in\Cat D(d,d')$, the following
diagram commutes.
\begin{equation*}
  \begin{tikzpicture}[->, >=stealth]
    \node (1) {$F(\Delta(j),d)$};
    \node (2) [right of=1, node distance=24mm] {$H(d)$};
    \node (3) [below of=1, node distance=12mm] {$F(\Delta(j'),d')$};
    \node (4) [below of=2, node distance=12mm] {$H(d')$};
    \tikzstyle{every node}=[midway,auto,font=\scriptsize]
    \draw (1) -- node {$\delta_{j,d}$} (2);
    \draw (1) -- node [swap] {$F(\Delta(\phi),g)$} (3);
    \draw (2) -- node {$H(g)$} (4);
    \draw (3) -- node {$\delta_{j',d'}$} (4);
  \end{tikzpicture}  
\end{equation*}
this results from the definition of $\Funcat{\Cat D}{\Cat E}$ and $F'$.

By our assumption on $F$, for each object $d$ of $\Cat D$ the
$J$-cocone $F(\gamma,d):F(\Delta,d)\Tonatural F(c,d)$ is colimiting in
$\Cat E$ and hence there is exactly one morphism
$\theta_d\in\Cat E(F(c,d),H(d))$ such that,
\begin{align}\label{eq:funcat-theta-def}
\forall j\in\Obj J\quad \theta_d\Compl F(\gamma_j,d)=\delta_{j,d}\,.
\end{align}
We prove that
$\theta=(\theta_d)_{d\in\Obj{\Cat D}}$ is a natural transformation
$F'(c)\Tonatural H$ so let $g\in\Cat D(d,d')$, we must prove that the
following diagram commutes.
\begin{equation*}
  \begin{tikzpicture}[->, >=stealth]
    \node (1) {$F(c,d)$};
    \node (2) [right of=1, node distance=24mm] {$H(d)$};
    \node (3) [below of=1, node distance=12mm] {$F(c,d')$};
    \node (4) [below of=2, node distance=12mm] {$H(d')$};
    \tikzstyle{every node}=[midway,auto,font=\scriptsize]
    \draw (1) -- node {$\theta_d$} (2);
    \draw (1) -- node [swap] {$F(c,g)$} (3);
    \draw (2) -- node {$H(g)$} (4);
    \draw (3) -- node {$\theta_{d'}$} (4);
  \end{tikzpicture}  
\end{equation*}
For any $j\in\Obj J$, we have
\begin{align*}
  H(g)\Compl\theta_d\Compl F(\gamma_j,d)
  &=H(g)\Compl\delta_{j,d}\quad\text{by definition of }\theta\\
  &=\delta_{j,d'}\Compl F(\Delta(j),g)\quad\text{by naturality of }\delta\\
  &=\theta_{d'}\Compl F(\gamma_j,d')\Compl F(\Delta(j),g)\\
  &=\theta_{d'}\Compl F(c,g)\Compl F(\gamma_j,d)
\end{align*}
and the required commutation follows by the uniqueness part of
universality from the fact that the cocone $F(\gamma,d)$ is colimiting.

This shows that $\theta\in\Funcat{\Cat D}{\Cat E}(F'(c),H)$. It
follows from~\Eqref{eq:funcat-theta-def} that for any $j\in\Obj J$,
one has $\theta\Compl F'(\gamma_j)=\delta$. Uniqueness follows from
the fact that any $\eta\in\Funcat{\Cat D}{\Cat E}(F'(c),H)$ such that
$\eta\Compl F'(\gamma_j)=\delta$ must satisfy the analogue
of~\Eqref{eq:funcat-theta-def} for each given $d\in\Obj D$ and hence
must be equal to $\theta$.
\Endproof

\subsection{Proof of Lemma~\ref{lemma:dense-nat-trans-extension}}
\Beginproof
Let $c\in\Cat C$, for each $(x,f)\in\Obj{\Slice Ic}$ (so that
$f\in\Cat C(I(x),c)$) we define
$\delta_{(x,f)}:(F\Compl\Slicepr c){(x,f)}=(F\Compl I)(x)\to G(c)$ by
\[
  \delta_{(x,f)}=G(f)\Compl\tau_x
\]
(remember indeed that
$\tau_x\in\Cat C((F\Compl I)(x),(G\Compl I)(x))$).

Then $\delta$ is a cocone $F\Compl\Slicepr c\Tonatural G(c)$ because,
given $t\in\Slice Ic((x,f),(y,g))$, we have
\begin{align*}
  \delta_{(y,g)}\Compl(F\Compl I)(t)
  &=G(g)\Compl\tau_{y}\Compl(FI(t))\quad\text{by definition of $\delta$}\\
  &=G(g)\Compl(G\Compl I)(t)\Compl\tau_x\quad\text{by naturality of }\tau\\
  &=G(g\Compl I(t))\Compl\tau_x\quad\text{by functoriality of }G\\
  &=G(f)\Compl\tau_x\quad\text{since }t\in \Slice Ic((x,f),(y,g))\,.
\end{align*}
Since the cocone $F\Slicec c:F\Compl\Slicepr c\Tonatural F(c)$ is
colimiting, it follows that there is exactly one morphism
$\Densext\tau_c\in\Cat C(F(c),G(c))$ such that, for each
$(x,f)\in\Obj{\Slice Ic}$, one has
$\delta_{(x,f)}=\Densext\tau_c\Compl F(\Slicec c_{(x,f)})$ that is
(coming back to the definitions of $\delta$ and $\Slicec c$), the
following diagram commutes
\begin{equation*}
  \begin{tikzpicture}[->, >=stealth]
    \node (1) {$F(I(x))$};
    \node (2) [right of=1, node distance=24mm] {$G(I(x))$};
    \node (3) [below of=1, node distance=12mm] {$F(c)$};
    \node (4) [below of=2, node distance=12mm] {$G(c)$};
    \tikzstyle{every node}=[midway,auto,font=\scriptsize]
    \draw (1) -- node {$\tau_x$} (2);
    \draw (1) -- node [swap] {$F(f)$} (3);
    \draw (2) -- node {$G(f)$} (4);
    \draw (3) -- node {$\Densext\tau_{c}$} (4);
  \end{tikzpicture}  
\end{equation*}
Notice that the uniqueness of this morphism implies, in the case
$c=I(x)$ and $f=\Id$, that $\Densext\tau_x=\tau_x$, so, for the first
statement of the theorem, we are left with proving that
$\Densext\tau_c$ is natural in $c$. So let $h\in\Cat C(c,c')$ and let
us prove that $\Densext\tau_{c'}\Compl
F(h)=G(h)\Compl\Densext\tau_c$. So let $f\in\Cat C(I(x),c)$, we have
\begin{align*}
  \Densext\tau_{c'}\Compl F(h)\Compl F(f)
  &= \Densext\tau_{c'}\Compl F(h\Compl f)\\
  &= G(h\Compl f)\Compl\tau_x\quad\text{by definition of }\Densext\tau\\
  &= G(h)\Compl G(f)\Compl\tau_x\\
  &= G(h)\Compl\Densext\tau_c\Compl F(f)
\end{align*}
and we obtain the expected commutation by
Lemma~\ref{lemma:dense-unique-morphism} and the fact that $F\Slicec c$
is colimiting.

As to the second part of the lemma, assume that $\tau$ is a natural
isomorphism whose inverse is $\sigma$, and that $G$ is also
cocontinuous, we get a unique natural transformation
$\Densext\sigma:G\Tonatural F$ such that
$\Densext\sigma\Compl I=\sigma$. Now
$\Densext\tau\Compl\Densext\sigma:G\Tonatural G$ satisfies
$(\Densext\tau\Compl\Densext\sigma)\Compl I=\tau\Compl\sigma=\Id$ and
hence by the uniqueness (applied to that natuarl transformation
$\Id:G\Compl I\Tonatural G\Compl I$) we get
$\Densext\tau\Compl\Densext\sigma=\Id$ and similarly
$\Densext\sigma\Compl\Densext\tau=\Id$ as contended.
\Endproof

\subsection{Proof of Lemma~\ref{lemma:dense-unique-morphism-multi}}
\Beginproof
By induction on $n$, the base case $n=0$ being trivial.  So for
$i=1,\dots,n$ let $I_i:\Cat C_i^0\to\Cat C_i$ be dense functors and
let $I:\Cat C^0\to\Cat C$ be a dense functor. Let
$F:\Cat C\times\prod_{i=1}^n\Cat C_i\to\Cat D$ be a separately
cocontinuous functor.

Given $c\in\Obj C$, we use $F_c:\prod_{i=1}^n\Cat C_i\to\Cat D$
for the functors obtained by fixing the first argument to $c$, notice
that $F_c$ is separately cocontinuous.

Let $c\in\Objc C$, $\Vect c\in\Obj{\prod_{i=1}^n\Cat C_i}$,
$d\in\Obj D$ and let $l,l'\in\Cat D(F(c,\Vect c),d)$ be such that for
all $x\in\Obj{\Cat C^0}$, $\Vect x\in\Obj{\prod_{i=1}^n\Cat C^0_i}$
and all $f\in\Cat C(I(x),c)$ and
$\Vect f\in\prod_{i=1}^n\Cat C_i(I_i(x_i),c_i)$ one has
$l\Compl F(f,\Vect f)=l'\Compl F(f,\Vect f)$.

To prove that $l=l'$ it suffices, by inductive hypothesis applied to
the functor $F_c$, to prove that for all
$\Vect x\in\Obj{\prod_{i=1}^n\Cat C^0_i}$ and
$\Vect f\in\prod_{i=1}^n\Cat C_i(I_i(x_i),c_i)$ one has
$l\Compl F(c,\Vect f)=l'\Compl F(c,\Vect f)$.  Let $k$ be the first of
these morphisms and $k'$ be the second one (with $\Vect x$ and
$\Vect f$ as above). By Lemma~\ref{lemma:dense-unique-morphism}
applied to the cocontinuous functor
$F(\Dummy,I_1(x_1),\dots,I_n(x_n))$ it suffices to show that for any
$x\in\Obj{\Cat C^0}$ and $f\in\Cat C(I(x),c)$, one has
$k\Compl F(f,I_1(x_1),\dots,I_n(x_n))=k'\Compl
F(f,I_1(x_1),\dots,I_n(x_n))$ which results from our assumption on $l$
and $l'$ and functoriality of $F$.
\Endproof

\subsection{Proof of Theorem~\ref{th:dense-concont-tnat-extension}}
\Beginproof
By induction on $n$, the base case being trivial. So for $i=1,\dots,n$
let $I_i:\Cat C_i^0\to\Cat C_i$ be dense functors and let
$I:\Cat C^0\to\Cat C$ be a dense functor. Let
$F,G:\Cat C\times\prod_{i=1}^n\Cat C_i\to\Cat D$ be functors and
assume that $F$ is separately cocontinuous.

For each $x\in\Obj C^0$, we
define a natural transformation
\[
  \tau(x):F_{I(x)}\Compl(\prod_{i=1}^nI_i)\Tonatural
  G_{I(x)}\Compl(\prod_{i=1}^nI_i)
\]
by setting $\tau(x)_{\Vect x}=\tau_{x,\Vect x}$. By inductive
hypothesis, there is an unique natural transformation
$\Densext{\tau(x)}:F_{I(x)}\Tonatural G_{I(x)}$ such that
$\Densext{\tau(x)}\Compl(\prod_{i=1}^nI_i)=\tau(x)$. So for each
$x\in\Obj{\Cat C^0}$, we have defined a morphism
$\Densext{\tau(x)}\in\Funcat{F'(I(x))}{G'(I(x))}$, we prove now that
it is natural in $x$.

Let $t\in\Cat C^0(x,y)$ and let
$\Vect c\in\Obj{\prod_{i=1}^n\Cat C_i}$, we must prove that the
following diagram commutes
\begin{equation*}
  \begin{tikzpicture}[->, >=stealth]
    \node (1) {$F(I(x),\Vect c)$};
    \node (2) [right of=1, node distance=24mm] {$G(I(x),\Vect c)$};
    \node (3) [below of=1, node distance=12mm] {$F(I(y),\Vect c)$};
    \node (4) [below of=2, node distance=12mm] {$G(I(y),\Vect c)$};
    \tikzstyle{every node}=[midway,auto,font=\scriptsize]
    \draw (1) -- node {$\Densext{\tau(x)}_{\Vect c}$} (2);
    \draw (1) -- node [swap] {$F(I(t),\Vect c)$} (3);
    \draw (2) -- node {$G(I(t),\Vect c)$} (4);
    \draw (3) -- node {$\Densext{\tau(y)}_{\Vect c}$} (4);
  \end{tikzpicture}  
\end{equation*}
Let
$\Vect x\in\Obj{\prod_{i=1}^n\Cat C^0_i}$ and
$\Vect f\in\prod_{i=1}^n\Cat C_i(I_i(x_i),c_i)$, we have
\begin{align*}
  G(I(t)&,\Vect c)\Compl \Densext{\tau(x)}_{\Vect c}\Compl F(I(x),\Vect f)\\
  &= G(I(t),\Vect c)\Compl G(I(x),\Vect f)
    \Compl\Densext{\tau(x)}_{I_1(x_1),\dots,I_n(x_n)}\\
  &\hspace{2cm}\text{by naturality of }\Densext{\tau(x)}\\
  &= G(I(t),\Vect f)\Compl\tau_{x,\Vect x}
    \quad\text{by ind.~hyp.~applied to }\tau(x)\\
  &= G(I(y),\Vect f)\Compl G(I(t),\Vect x)\Compl\tau_{x,\Vect x}
  \quad\text{func.~of }G\\
  &= G(I(y),\Vect f)\Compl \tau_{y,\Vect x}\Compl F(I(t),\Vect x)
  \quad\text{nat.~of }\tau\\
  &= \Densext{\tau(y)}_{\Vect c}\Compl F(I(y),\Compl f)\Compl F(I(t),\Vect x)
  \quad\text{by ind.~hyp.~for }\tau(y)\\
  &= \Densext{\tau(y)}_{\Vect c}\Compl F(I(t),\Vect c)\Compl F(I(x),\Vect f)
\end{align*}
and hence by Lemma~\ref{lemma:dense-unique-morphism-multi},
$G(I(t),\Vect c)\Compl \Densext{\tau(x)}_{\Vect c}
=\Densext{\tau(y)}_{\Vect c}\Compl F(I(t),\Vect c)$
as contended.

Let $\rho:F'\Compl I\Tonatural G'\Compl I$ be defined by
$\rho_x=\Densext{\tau(x)}$. Since $F'$ is cocontinuous by
Lemma~\ref{lemma:functor-transpose-cocont}, we know
by Lemma~\ref{lemma:dense-nat-trans-extension} that there is
exactly one $\Densext\rho$ such that $\Densext\rho\Compl I=\rho$.

We set $\Densext\tau_{c,\Vect c}=(\Densext\rho_c)_{\Vect c}$, this
family of morphisms $\Densext\tau$ is a natural transformation
$F\Tonatural G$ such that
$\Densext\tau\Compl(I\times\prod_{i=1}^n I_i)=\tau$.

Uniqueness is straightforward: assume $\theta:F\Tonatural G$ satisfies
$\theta\Compl(I\times\prod_{i=1}^n I_i)=\tau$. Then
$\theta_{I(x),\Vect I(\Vect x)}=\tau(x)_{\Vect x}$ and hence by the
uniqueness of $\Densext{\tau(x)}$ we must have
$\theta_{I(x),\Vect c}=\Densext{\tau(x)}_{\Vect c}=(\rho_x)_{\Vect
  c}$. Therefore, the natural transformation $\theta':F'\Tonatural G'$
defined by $(\theta'_c)_{\Vect c}=\theta_{c,\Vect c}$ satisfies
$\theta'\Compl I=\rho$ from which it follows that
$\theta'=\Densext\rho$, that is $\theta=\Densext\tau$.

The last statement of the lemma is proven exactly as the last
statement of Lemma~\ref{lemma:dense-nat-trans-extension}.
\Endproof

\subsection{Proof of Lemma~\ref{lemma:commutation-cone-sums}}
\Beginproof
Let $K\subseteq I\times J$ be finite and les $K_1\subseteq I$ and
$K_2\subseteq J$ be its projections,
$\Norm{\sum_{(i,j)\in K}x_{i,j}}\leq\Norm{\sum_{i\in K_1}\sum_{j\in
    K_2}x_{i,j}}$ by monotonicity of the norm. So
$\Norm{\sum_{(i,j)\in K}x_{i,j}}\leq\Norm{\sum_{i\in K_1}\sum_{j\in
    J}x_{i,j}}$ and hence the family
$\Norm{\sum_{(i,j)\in K}x_{i,j}}_{K\in\Partfin{I\times J}}$ is bounded
by our assumption that $(\sum_{j\in J}x_{i,j})_{i\in I}$ is
summable. The stated equations result from continuity of addition.
\Endproof

\subsection{Proof of Lemma~\ref{lemma:pcoh-closed-charact}}
\Beginproof
The $\Implies$ implication is easy (see~\cite{DanosEhrhard08}), we
prove the converse, which uses the Hahn-Banach theorem in finite
dimension.  Let $v\in\Realpto I$ be such that $v\notin\cU$. We must prove
that there exists $u'\in\Orth\cU$ such that $\Eval v{u'}>1$ and
$\forall u\in\cU\,\Eval u{u'}\leq 1$. Given $J\subseteq I$ and
$w\in\Realpto I$, let $\FamRestr wJ$ be the element of $\Realpto I$
which takes value $w_j$ for $j\in J$ and $0$ for $j\notin J$. Then $v$
is the lub of the increasing sequence
$\{\FamRestr v{\{\List i1n\}}\St n\in\Nat\}$ (where $i_1,i_2,\dots$ is
any enumeration of $I$) and hence there must be some $n\in\Nat$ such
that $\FamRestr v{\{\List i1n\}}\notin \cU$.  Therefore it suffices to
prove the result for $I$ finite, what we assume now.  Let
$\cG=\{u\in\Realto I\St (\Absval{u_i})_{i\in I}\in \cU\}$ which is a
convex subset of $\Realto I$.  Let
$\lambda_0=\sup\{\lambda\in\Realp\St \lambda v\in \cU\}$. By our
closeness assumption on $\cU$, we have $\lambda_0v\in\cU$ and
therefore $\lambda_0<1$. Let $h:\Real v\to\Real$ be defined by
$h(\lambda v)=\lambda/\lambda_0$ ($\lambda_0\not=0$ by our assumptions
about $\cU$ and because $I$ is finite). Let $q:\Realto I\to\Realp$ be
the gauge of $\cG$, which is the semi-norm given by
\(
q(w)=\inf\{\epsilon >0\St w\in\epsilon \cG\}
\).
It is actually a norm by our assumptions on $\cU$. Observe that
$h(w)\leq q(w)$ for all $w\in\Real v$: this boils down to showing that
$\lambda\leq \lambda_0 q(\lambda v)=\Absval\lambda \lambda_0q(v)$ for
all $\lambda\in\Real$ which is clear since $\lambda_0q(v)=1$ by
definition of these numbers.  Hence, by the Hahn-Banach Theorem, there
exists a linear $l:\Realto I\to\Real$ such that $\Absval l\leq q$ and
which coincides with $h$ on $\Real v$. Let $v'\in\Realto I$ be such
that $\Eval w{v'}=l(w)$ for all $w\in\Realto I$ (using again the
finiteness of $I$). Let $u'\in\Realpto I$ be defined by
$u'_i=\Absval{v'_i}$. It is clear that $\Eval v{u'}>1$: since
$v\in\Realpto I$ we have
$\Eval v{u'}\geq\Eval v{v'}=l(v)=h(v)=1/\lambda_0>1$. Let
$N=\{i\in I\St v'_i<0\}$.  Given $w\in \cU$, let $\bar w\in\Realto I$
be given by $\bar w_i=-w_i$ if $i\in N$ and $\bar w_i=w_i$
otherwise. Then $\Eval w{v'}=\Eval{\bar w}{u'}=l(\bar w)\leq 1$ since
$\bar w\in \cG$ (by definition of $\cG$ and because $w\in \cU$). It
follows that $u'\in\Orth\cU$.
\Endproof

\subsection{Proof of Lemma~\ref{lemma:pcoh-as-closures}}
\Beginproof
Let $\cG\subseteq\Realpto I$. Let $\Cvx\cG$ be the set of all the
elements of $\Realpto I$ which are of shape
$\sum_{j=1}^k\alpha_j u(j)$ where $u(j)\in\cG$ and
$\sum_{j=1}^k\alpha_j=1$.  We use $\cG^+$ for the set of all
$u\in\Realpto I$ such that there is a monotone sequence
$(u(n))_{n\in\Nat}$ of elements of $\Cvx\cG$ such that
$u\leq\sup_{n\in\Nat}u(n)$. Clearly $\cG\subseteq\cG^+$. For each
ordinal $\beta$, we define $\cU(\beta)\subseteq\Realpto I$ by
induction as follows: $\cU(0)=\cU$, $\cU(\beta+1)=\cU(\beta)^+$ and,
if $\beta$ is limit and $>0$, then
$\cU(\beta)=\cup_{\gamma<\beta}\cU(\gamma)$. This sequence is clearly
monotone for $\subseteq$. Let $\beta$ be the least ordinal number such
that $\cU(\beta+1)=\cU(\beta)$. We have
$\Biorth\cU=\cU(\beta)$ since $\cU(\beta)$ is the least subset of
$\Realpto I$ which contains $\cU$, is convex, downwards-closed and
closed under the lubs of monotone sequences, and therefore satisfies
$\Biorth{\cU(\beta)}=\cU(\beta)$ by
Lemma~\ref{lemma:pcoh-closed-charact}.

To prove our contention, il suffices therefore to prove that, for any
$\cG\subseteq\Realpto I$ and any $h:I\to P$ such that
$\forall u\in\cG\ \sum_{a\in I}u_ah(a)\in\Cuball P$, one has
$\forall u\in\cG^+\ \sum_{a\in I}u_ah(a)\in\Cuball P$, the result will
follow by ordinal induction. So assume that $\cG$ and $h$ satisfy
these hypotheses. First let $v\in\Cvx\cG$, say
$v=\sum_{j=1}^k\alpha_j v(j)$ where $v(j)\in\cG$ and
$\alpha_j\in\Realp$ such that $\sum_{j=1}^k\alpha_j=1$. Then
\begin{align*}
  \sum_{a\in I}v_ah(a)
  &=\sum_{a\in I}(\sum_{j=1}^k\alpha_jv(j)_a)h(a)\\
  &=\sum_{j=1}^k\alpha_j(\sum_{a\in I}v(j)_ah(a))\in\Cuball P
\end{align*}
by convexity\footnote{And actually also
  closeness because this computation uses implicitely restrictions of
  the sum over $I$ to finite subsets of $I$.} of $\Cuball P$.

Let now $u\in\cG^+$ and let $(u(n))_{n\in\Nat}$ be a monotone sequence
in $\Cvx\cG$ such that $u\leq\sup_{n\in\Nat}u(n)$. For each $n$ we
have $\sum_{a\in I}u(n)_ah(a)\in\Cuball P$ by what we have just proven
and hence $\sup_{n\in\Nat}\sum_{a\in I}u(n)_ah(a)\in\Cuball P$ by
completeness of $P$ since the sequence
$(\sum_{a\in I}u(n)_ah(a))_{n\in\Nat}$ is monotone. By continuity of
the algebraic operations in $P$ we have
$\sup_{n\in\Nat}\sum_{a\in I}u(n)_ah(a)=\sum_{a\in
  I}\sup_{n\in\Nat}u(n)_ah(a)$ and since
$\forall a\in I\ u_a\leq\sup_{n\in\Nat}u(n)_a$ we get
$\sum_{a\in I}u_ah(a)\in\Cuball P$ as contended.

The fact that $\bar h\in\SCCLin(\ConeofPCS(I,\Biorth\cU),P)$ results
clearly from its definition and from the fact that it maps
$\Biorth\cU$ to $\Cuball P$.
\Endproof

\subsection{Proof of Lemma~\ref{lemma:limpl-continuous}}
\Beginproof
It suffices to check that it preserves all small products and binary
equalizers. Let first $\Vect Q=(Q_i)_{i\in I}$ be a family of objects
of $\SCCLin$. Any element of $\Limpl P{\prod\Vect Q}$ is of shape
$\Tuple{f_i}_{i\in I}$ with $f_i\in\Limpl P{Q_i}$ for each $i$ and
this defines a map
$\theta_{\Vect Q}:\Limplp{P}{\prod\Vect Q}\to\prod_{i\in I}\Limplp
P{Q_i}$ which is a bijection. This map is linear and continuous
because all operations are calculated pointwise (wrt.~the argument of
functions) and componentwise (in the product indexed by $I$). The fact
that $\Norm{\theta_{\Vect Q}}=1$ results from the fact that all the
norms involved are computed as lubs in $\Realp$. To check that
$\theta_{\Vect Q}$ is an iso it suffices to check that
$\Funinv{\theta_{\Vect Q}}$ is continuous. Let us check this point:
let $(f(n))_{n\in\Nat}$ be a non-decreasing sequence in
$\Cuballp{\prod_{i\in I}\Limplp P{Q_i}}$ so that
$f(n)=(f(n)_i)_{i\in I}$, where $f(n)_i\in\Limpl P{Q_i}$ and for each
$i\in I$ the sequence of functions $(f(n)_i)_{n\in\Nat}$ is
non-decreasing, and for each $x\in\Cuball P$, one has
$\forall n\in\Nat\ \Norm{f(n)_i(x)}\leq 1$. Then
$f=\sup_n f(n)\in\Cuballp{\prod_{i\in I}\Limplp P{Q_i}}$ is
characterized by $f(x)_i=\sup_{n\in\Nat}f(n)_i(x)$. On the other hand,
$g=\Funinv{\theta_{\Vect Q}}(f)\in\Limpl P{\prod\Vect Q}$ is given by
$g(x)=(f(x)_i)_{i\in I}$ so that $g(x)$ is the lub in $\prod\Vect Q$
of the sequence $(f(n)(x))_{n\in\Nat}$ and since lubs of sequences of
functions are computed pointwise, this proves our contention. So
$\theta_{\Vect Q}$ is an iso in $\SCCLin$ and its naturality is
obvious.

Next consider $f_1,f_2\in\SCCLin(Q,R)$ and let $(E,e)$ be the
corresponding equalizer ($E$ is the cone of elements $x$ of $Q$ such
that $f_1(x)=f_2(x)$ and $e:E\to Q$ is the inclusion). Then
$\Limpl P{f_i}\in\SCCLin(\Limpl P{Q},\Limpl P{R})$ (for $i=1,2$) maps
$h$ to ${f_i}\Compl {h}$. The equalizer of these two maps is the cone
of all $h\in\Limpl PQ$ such that $f_1\Compl h=f_2\Compl h$, that is
$\forall x\in P\ f_1(h(x))=f_2(h(x))$, equivalently $h\in\Limpl
PE$. And the inclusion map $\Limplp PE\to\Limplp PQ$ is equal to
$\Limpl Pe$. Hence the equalizer of $\Limpl P{f_1}$ and
$\Limpl P{f_2}$ is $(\Limpl PE,\Limpl Pe)$ which proves that the
functor $\Limpl P\Wcard$ preserves equalizers, and hence preserves all
small limits.
\Endproof

\subsection{Proof of Lemma~\ref{lemma:func-staboflin-continuous}}
\Beginproof
It suffices to prove that $\Staboflin$ preserves small products and
binary equalizers. The first statement results from the fact that
$\SCCStab$ is cartesian with products defined as in $\SCCLin$. Let us
prove the second one so let $f_1,f_2\in\SCCLin(P,Q)$ and $(E,e)$ be
the corresponding equalizer in $\SCCLin$ (that is
$E=\Eset{x\in P\St f_1(x)=f_2(x)}$ and $e:E\to P$ is the obvious
inclusion, see the proof of Theorem~\ref{th:cones-lin-complete}). We
prove that $(E,e)$ is the equalizer of $f_1$ and $f_2$ in $\SCCStab$
so let $g\in\SCCStab(R,P)$ be such that $f_1\Comp g=f_2\Comp g$, that
is $\forall z\in\Cuball R\ g(z)\in E$. Let $h:\Cuball R\to E$ be
defined by $h(z)=g(z)$, then $h$ is stable because $g$ is  and
$E$ inherits its structure from $P$ (which also entails that
$h(\Cuball R)\subseteq\Cuball E$ since
$g(\Cuball R)\subseteq\Cuball P$).  And $h$ is the unique element of
$\SCCStab(R,E)$ such that $g=e\Comp h$ which proves our contention.
\Endproof

\subsection{Proof of Lemma~\ref{lemma:tens-measurable}}
\Beginproof
Let $f\in\SCCLinm(\Tensc{P}{Q},R)$ and $g=\Curlin(f)$, we prove that
$g\in\SCCLinm(P,\Limplm{Q}{R})$. Let first $x\in\Mconec P$, we prove
that $g(x)\in\Mconec{\Limplm QR}$ so let $\delta\in\Mconet Q_p$, we
prove that $g(x)\Comp\delta\in\Mpath R_p$. Let $m\in\Mconet Q_q$, we have
\begin{align*}
\Metabs{(r,s)}{m(s)(g(x)(\delta(r)))} =\Metabs{(r,s)}{m(s)(f(\Tensc
  x{\delta(r)}))}\,.
\end{align*}
Let $\gamma=\Metabs wx\in\Mconet P_\Rzero$ we
have $f\Comp{\Tenscp\gamma\delta}\in\Mpath R_{p}$ by
Lemma~\ref{lemma:tens-of-paths} and by our assumption about $f$ and
hence
$\Metabs{(r,s)}{m(s)(g(x)(\delta(r)))}\in\MSP(\Mref(p+q),\Realp)$
so that $g(x)\in\Limplm QR$. We prove that
$g\in\SCCLinm(P,\Limplm QR)$ so let $\gamma\in\Mpath P_p$ and let us
show that $g\Comp\gamma\in\Mpath{\Limplm QR}_p$; applying the
definition of $\Limplm QR$, let $\delta\in\Mpath Q_q$ and
$l\in\Mconet R_q$,
we have
\begin{align*}
  \Metabs{(r,s)}{(\Lftest\delta
  l)(s)(g(\gamma(r))}
  &=\Metabs{(r,s)}{l(s)(g(\gamma(r))(\delta(s)))}\\
  &=\Metabs{(r,s)}{l(s)(f(\Tensc{\gamma(r)}{\delta(s)}))} 
\end{align*}
and we know by our assumtion on $f$ and by
Lemma~\ref{lemma:tens-of-paths} that
$f\Comp(\Tensc\gamma\delta)\in\Mpath{R}_{p+q})$ and hence
\[
  \Metabs{(r,s,s')}{l(s')(f(\Tensc{\gamma(r)}{\delta(s)}))}
  \in\MSP(\Mref(p+q+q),\Realp)
\]
from which
$\Metabs{(r,s)}{(\Lftest\delta
  l)(s)(g(\gamma(r))}\in\MSP(\Mref(p+q),\Realp)$ follows since
$\MREFC$ is cartesian and measurability tests are closed under
precomposition by morphisms of $\MREFC$.

Conversely, let $g\in\SCCLinm(P,\Limplm QR)$ and let
$f=\Funinv\Curlin(g)\in\SCCLin(\Tensc{\Mconec P}{\Mconec Q},\Mconec
R)$ so that $f$ is uniquely characterized by the fact that
$f(\Tensc xy)=g(x)(y)$ for all $x\in\Mconec P$ and $y\in\Mconec Q$. We
must prove that $f\in\SCCLinm(\Tensc PQ,R)$ so let
$\theta\in\Mpath{\Tensc PQ}_{p_0}$, we must show that
$f\Comp\theta\in\Mpath R_{p_0}$. Let $l\in\Mconet R_k$ and let us
prove that
$\Metabs{(r_0,w)}{l(w)(f(\theta(r_0))}\in\MSP(\Mref(p_0+k),\Realp)$. For
each $w\in\Mref(k)$, we have
$\Metabs{z}{l(w)(f(z))}\in\Cdual{\Tenscp{\Mconec P}{\Mconec Q}}$
because $f$ is linear and continuous and $l(w)\in\Cdual{R}$.  So let
$m\in\Cdual{\Tenscp{\Mconec P}{\Mconec Q}}^{\Mref(k)}$ be defined as
$m=\Metabs{w}{\Metabs{z}{l(w)(f(z))}}$, we claim that
$m\in\Mconet{\Tensc PQ}_k$. The first condition (namely for all
$z\in\Tensc{\Mconec P}{\Mconec Q}$, one has
$\Metabs w{m(w)(z)}\in\MSP(\Mref(k),\Realp)$) being obviously
satisfied, we check the second one so let $\gamma\in\Mpath P_p$ and
$\delta\in\Mpath Q_q$.
We have
\begin{align*}
  \Metabs{(r,s,w)}{m(w)(\Tensc{\gamma(r)}{\delta(s)})}
  &=\Metabs{(r,s,w)}{l(w)(f(\Tensc{\gamma(r)}{\delta(s)}))}\\
  &=\Metabs{(r,s,w)}{l(w)(g(\gamma(r))(\delta(s)))}\,.
\end{align*}
We set
\begin{align*}
  \delta'&=\Metabs{(s,w)}{\delta(s)}:\Mref(q+k)\to\Mconec Q\\
  l'&=\Metabs{(s,w)}{\Metabs z{l(w)(z)}}\in{\Cdual R}^{\Mref(q+k)}\,.
\end{align*}
Then\footnote{Because $\MREFC$ is cartesian and measurability tests
  and paths are closed under precomposition by morphisms of $\MREFC$.}
$\delta'\in\Mpath Q_{q+k}$ and $l'\in\Mconet R_{q+k}$ and therefore
$\Lftest{\delta'}{l'}\in\Mconet{\Limplm QR}_{q+k}$.  We know that
$g\Comp\gamma\in\Mpath{\Limplm QR}_p$ and hence
$\Metabs{(r,s,w)}{(\Lftest{\delta'} {l'})(s,w)(g(\gamma(r)))}$ is
measurable $\Mref(p+q+k)\to\Realp$. Now observe that
{\small
  \begin{align*}
  \Metabs{(r,s,w)}{(\Lftest{\delta'}{l'})(s,w)(g(\gamma(r)))}
  &= \Metabs{(r,s,w)}{l(w)(g(\gamma(r))(\delta(s)))}
  \end{align*}
}%
so we have proven that $m\in\Mconet{\Tensc PQ}_k$. But remember that
$\theta\in\Mpath{\Tensc PQ}_{p_0}$, we have therefore
$\Metabs{(r_0,w)}{m(w)(\theta(r_0))}\in\MSP(\Mref(p_0+k),\Realp)$ and
since $m(w)(\theta(r_0))$ is nothing but $l(w)(f(\theta(r_0)))$ we
have $f\Comp\theta\in\Mpath R_{p_0}$.
\Endproof

\subsection{Proof of Theorem~\ref{lemma:exp-measurable}}
\Beginproof
Let $f\in\SCCLinm(\Exclc P,Q)$, the associated
$g\in\SCCStab(\Mconec P,\Mconec Q)$ is defined by $g(x)=f(\Promc x)$.
Let $\gamma\in\Mpath P_p$, we have
$g\Comp\gamma=f\Comp\Promc\gamma\in\Mpath Q_p$ since
$\Promc\gamma\in\Mpath{\Exclc P}_p$ and hence $g\in\SCCStab(P,Q)$. Now
let $g\in\SCCStabm(P,Q)$ and let
$f\in\SCCLin(\Exclc{\Mconec P},\Mconec Q)$ be the associated linear
map, uniquely characterized by
$\forall x\in\Cuball P\ g(x)=f(\Promc x)$. Let
$\theta\in\Mpath{\Exclc P}_p$, we prove that
$f\Comp\theta\in\Mpath Q_p$ so let $m\in\Mconet Q_q$, we define
$l=\Metabs{s}{\Metabs z{m(s)(f(z))}}\in\Cdual{\Exclc{\Mconec
    P}}^{\Mref(q)}$ (linearity and continuity of $l(s)$ follows from
those of $f$). The fact that
$\Metabs s{l(s)(z)}\in\MSP(\Mref(q),\Realp)$ for each
$z\in\Exclc{\Mconec P}$ follows from $m\in\Mconet Q_q$. Let
$\gamma\in\Mpath P_{p_0}$, we have
\begin{align*}
  \Metabs{(r_0,s)}{l(s)(\Promc{\gamma(r_0)})}
  &=\Metabs{(r_0,s)}{m(s)(g(\gamma(r_0))}\\
  &\in\MSP(\Mref(r_0+s),\Realp) 
\end{align*}
since $g$ is measurable, hence $m\in\Mconet{\Exclc P}$. Since
$\theta\in\Mpath{\Exclc P}_p$ it follows that
$\Metabs{(r,s)}{l(s)(\theta(r))}\in\MSP(\Mref(r+s),\Realp)$ but
$l(s)(\theta(r))=m(s)(f(\theta(r)))$ and so we have proven that
$f\Comp\theta\in\Mpath Q_p$.
\Endproof

\subsection{The pentagon}\label{sec:pentagon}

We have to prove commutation of the external pentagon of
Figure~\ref{fig:pentagon} where the morphisms $\beta_i$, $\gamma_i$
are instances\footnote{Possibly involving tensorisations with
  identities, the same for the next uses of the word ``instance''.} of
$\Assocc$, $\alpha_i$ are obtained by applying $\ConeofPCS$ to
$\Assoc$ and $\pi_i$ are instances of $\Tensciso$, that is
$\beta_2\Compl\beta_1=\beta_5\Compl\beta_4\Compl\beta_3$. This is
reduced to the commutation of the internal pentagon involving
$\alpha_1,\dots,\alpha_5$ by observing that
\begin{align*}
  \beta_2\Compl\beta_1=\Funinv{(\pi_{11}\Compl\pi_{10}\Compl\pi_9)}
  \Compl\alpha_2\Compl\alpha_1\Compl(\pi_3\Compl\pi_2\Compl\pi_1)
\end{align*}
that is
$\pi_{11}\Compl\pi_{10}\Compl\pi_9
\Compl\beta_2\Compl\beta_1=\alpha_2\Compl\alpha_1\Compl(\pi_3\Compl\pi_2\Compl\pi_1)$
and similarly for $\beta_5\Compl\beta_4\Compl\beta_3$. This is done by
pasting five kinds of commutative squares of which we give examples,
explaining why they commute.

\begin{itemize}
\item The diagram involving $\beta_1$, $\pi_4$, $\pi_1$ and $\gamma_1$
  which commutes by naturality of $\Assocc$.
\item The diagram involving $\pi_4$, $\pi_6$, $\pi_5$ and $\pi_7$
  which commutes by functoriality of $\ITensc$.
\item The diagram involving $\gamma_1$, $\pi_6$, $\pi_8$, $\pi_2$,
  $\pi_3$ and $\alpha_1$ whose commutation results from the definition
  of $\Assocli$ and $\Assocc$.
\item The diagram involving $\beta_3$, $\pi_{13}$, $\pi_{14}$,
  $\pi_1$, $\pi_2$ and $\alpha_6$ whose commutation results from the
  definition of $\Assocli$ and $\Assocc$.
\item The diagram involving $\alpha_6$, $\pi_{12}$, $\pi_3$ and
  $\alpha_3$ which results from the naturality of $\Tensciso$.
\end{itemize}

\newcommand\Lab[1]{}

\begin{figure*}
  \centering
  \begin{tikzpicture}[->, >=stealth]
    \tikzstyle{every node}=[font=\small]
    \node (1) {$\Tensc{\Tenscp{\Tenscp{\ConeofPCS X_1}
          {\ConeofPCS X_2}}{\ConeofPCS X_3}}{\ConeofPCS X_4}$\Lab 1};
    \node (11) at (2,-2)
    {$\Tensc{\Tenscp{\ConeofPCS\Tensp{X_1}
          {X_2}}{\ConeofPCS X_3}}{\ConeofPCS X_4}$\Lab{11}};
    \node (12) at (2,-4.5)
    {$\Tensc{\ConeofPCS\Tensp{\Tensp{X_1}
          {X_2}}{X_3}}{\ConeofPCS X_4}$\Lab{12}};
    \node (131) at (2,-16)
    {$\Tensc{\ConeofPCS\Tensp{X_1}{\Tensp{X_2}
          {X_3}}}{\ConeofPCS X_4}$\Lab{131}};
    \node (132) at (4.5,-6.5)
    {$\ConeofPCS\Tensp{\Tensp{\Tensp{X_1}{X_2}}
          {X_3}}{X_4}$\Lab{132}};
    \node (2) [ right of=1, node distance=140mm ]
    {$\Tensc{\Tenscp{\ConeofPCS X_1}{\ConeofPCS X_2}}
      {\Tenscp{\ConeofPCS X_3}{\ConeofPCS X_4}}$\Lab 2};
    \node (24) at (11.5,-2.8)
    {$\Tensc{\Tenscp{\ConeofPCS X_1}{\ConeofPCS X_2}}
      {\ConeofPCS\Tensp{X_3}{X_4}}$\Lab{24}};
    \node (241) at (11.5,-7.5)
    {$\Tensc{\ConeofPCS X_1}{\Tenscp{\ConeofPCS X_2}
      {\ConeofPCS\Tensp{X_3}{X_4}}}$\Lab{241}};
    \node (21) at (9,-2)
    {$\Tensc{\ConeofPCS\Tensp{X_1}{X_2}}
      {\Tenscp{\ConeofPCS X_3}{\ConeofPCS X_4}}$\Lab{21}};
    \node (22) at (9,-4.5)
    {$\Tensc{\ConeofPCS\Tensp{X_1}{X_2}}
      {\ConeofPCS\Tensp{X_3}{X_4}}$\Lab{22}};
    \node (23) at (9,-6.5)
    {$\ConeofPCS\Tensp{\Tensp{X_1}{X_2}}
      {\Tensp{X_3}{X_4}}$\Lab{23}};
    \node (3) [ below of=1, node distance=200mm ]
    {$\Tensc{\Tenscp{\ConeofPCS X_1}{\Tenscp{\ConeofPCS X_2}
          {\ConeofPCS X_3}}}{\ConeofPCS X_4}$\Lab 3};
    \node (31) at (2,-18)
    {$\Tensc{\Tenscp{\ConeofPCS X_1}{\ConeofPCS\Tensp{X_2}
          {X_3}}}{\ConeofPCS X_4}$\Lab{31}};
    \node (4) [ below of=2, node distance=200mm ]
    {$\Tensc{\ConeofPCS X_1}{\Tenscp{\Tenscp{\ConeofPCS X_2}
          {\ConeofPCS X_3}}{\ConeofPCS X_4}}$\Lab 4};
    \node (41) at (11.5,-18)
    {$\Tensc{\ConeofPCS X_1}{\Tenscp{\ConeofPCS\Tensp{X_2}
          {X_3}}{\ConeofPCS X_4}}$\Lab{41}};
    \node (2bis) [ right of=2, node distance=30mm ] {};
    \node (5) [ below of=2, node distance=55mm ]
    {$\Tensc{\ConeofPCS X_1}{\Tenscp{\ConeofPCS X_2}
        {\Tenscp{\ConeofPCS X_3}{\ConeofPCS X_4}}}$\Lab 5};
    \node (51) at (9,-11.3)
    {$\ConeofPCS\Tensp{X_1}{\Tensp{X_2}{\Tensp{X_3}{X_4}}}$\Lab{51}};
    \node (511) at (11.5,-9.6)
    {$\Tenscp{\ConeofPCS X_1}{\ConeofPCS\Tensp{X_2}{\Tensp{X_3}{X_4}}}$\Lab{511}};
    \node (43) at (9,-14)
    {$\ConeofPCS\Tensp{X_1}{\Tensp{\Tensp{X_2}{X_3}}{X_4}}$\Lab{43}};
    \node (431) at (11.5,-16)
    {$\Tensc{\ConeofPCS X_1}{\ConeofPCS\Tensp{\Tensp{X_2}
          {X_3}}{X_4}}$\Lab{431}};
    \node (32) at (4.5,-14)
    {$\ConeofPCS\Tensp{\Tensp{X_1}{\Tensp{X_2}{X_3}}}{X_4}$\Lab{32}};
    \tikzstyle{every node}=[midway,auto,font=\scriptsize]
    \draw (1) -- node {$\beta_1$} (2);
    \draw (1) -- node {$\pi_1$} (11);
    \draw (11) -- node {$\pi_2$} (12);
    \draw (12) -- node [swap] {$\alpha_6$} (131);
    \draw (12) -- node {$\pi_3$} (132);
    \draw (2) -- node {$\pi_4$} (21);
    \draw (2) -- node {$\pi_5$} (24);
    \draw (11) -- node {$\gamma_1$} (21);
    \draw (21) -- node {$\pi_6$} (22);
    \draw (22) -- node {$\pi_8$} (23);
    \draw (24) -- node {$\pi_7$} (22);
    \draw (132) -- node {$\alpha_1$} (23);
    \draw (1) -- node [swap] {$\beta_3$} (3);
    \draw (3) -- node {$\beta_4$} (4);
    \draw (3) -- node {$\pi_{13}$} (31);
    \draw (31) -- node {$\pi_{14}$} (131);
    \draw (4) -- node [swap] {$\beta_5$} (5);
    \draw (2) -- node {$\beta_2$} (5);
    \draw (5) -- node {$\pi_9$} (241);
    \draw (24) -- node [swap] {$\gamma_2$} (241);
    \draw (23) -- node {$\alpha_2$} (51);
    \draw (241) -- node [swap] {$\pi_{10}$} (511);
    \draw (511) -- node {$\pi_{11}$} (51);
    \draw (4) -- node {$\pi_{15}$} (41);
    \draw (31) -- node {$\gamma_3$} (41);
    \draw (43) -- node [swap] {$\alpha_5$}(51);
    \draw (41) -- node {$\pi_{16}$} (431);
    \draw (431) -- node {$\pi_{17}$} (43);
    \draw (431) -- node {$\alpha_7$} (511);
    \draw (32) -- node {$\alpha_4$} (43);
    \draw (132) -- node [swap] {$\alpha_3$} (32);
    \draw (131) -- node {$\pi_{12}$} (32);
  \end{tikzpicture}
  \caption{Pentagon diagram}
  \label{fig:pentagon}
\end{figure*}



%

\end{document}